%% file: main.tex
\documentclass[times,onecolumn]{aastex62}

\usepackage{cases}
\usepackage{graphicx}
\usepackage{subfigure}
\usepackage{natbib}
\usepackage{color}
\usepackage{tabularx}
\usepackage{bm}
\usepackage{threeparttable}

\newcommand{\thetae}{\theta_{\rm E}}
\newcommand{\teff}{t_{\rm eff}}
\newcommand{\Icat}{I_{\rm cat}}
\newcommand{\pie}{\pi_{\rm E}}

\newcommand{\te}{t_{\rm E}}
\newcommand{\event}{OGLE-2019-BLG-1053}
\newcommand{\eventLb}{OGLE-2019-BLG-1053Lb}

\newcommand{\Sp}{{\it Spitzer}}

\DeclareGraphicsExtensions{.pdf,.png,.jpg}

\shorttitle{}
\shortauthors{Zang et al.}

\begin{document}
\title{{\large Systematic KMTNet Planetary Anomaly Search, Paper I: OGLE-2019-BLG-1053Lb, A Buried Terrestrial Planet}}

\correspondingauthor{Weicheng Zang}
\email{zangwc17@mails.tsinghua.edu.cn}

\input{author.tex}

\input{abstract}

\section{Introduction}\label{intro}
\input{intro}

\section{Anomaly Search}\label{anomaly}
\input{anomaly}

\section{Observations of \event}\label{obser}
\input{obser}

\section{Light-curve Analysis}\label{model}

\input{model}

\section{Lens Properties}\label{lens}
\input{lens}

\section{Discussion}\label{dis}

\input{dis}

\bibliography{Zang.bib}

\input{table.tex}

\input{figure.tex}

\end{document}

%% file: author.tex
\author[0000-0001-6000-3463]{Weicheng Zang}
\affiliation{Department of Astronomy, Tsinghua University, Beijing 100084, China}

\author{Kyu-Ha Hwang}
\affiliation{Korea Astronomy and Space Science Institute, Daejon 34055, Republic of Korea}

\author[0000-0001-5207-5619]{Andrzej Udalski}
\affiliation{Astronomical Observatory, University of Warsaw, Al. Ujazdowskie 4, 00-478 Warszawa, Poland}

\author{Tianshu Wang}
\affiliation{Department of Astrophysical Sciences, Princeton University}

\author[0000-0003-4027-4711]{Wei Zhu}
\affiliation{Department of Astronomy, Tsinghua University, Beijing 100084, China}

\author{Takahiro Sumi}
\affiliation{Department of Earth and Space Science, Graduate School of Science, Osaka University, Toyonaka, Osaka 560-0043, Japan}

\author{Jennifer C. Yee}
\affiliation{Center for Astrophysics $|$ Harvard \& Smithsonian, 60 Garden St.,Cambridge, MA 02138, USA}

\author{Andrew Gould}
\affiliation{Max-Planck-Institute for Astronomy, K\"onigstuhl 17, 69117 Heidelberg, Germany}
\affiliation{Department of Astronomy, Ohio State University, 140 W. 18th Ave., Columbus, OH 43210, USA}

\author{Shude Mao}
\affiliation{Department of Astronomy, Tsinghua University, Beijing 100084, China}
\affiliation{National Astronomical Observatories, Chinese Academy of Sciences, Beijing 100101, China}

\author{Xiangyu Zhang}
\affiliation{Department of Astronomy, Tsinghua University, Beijing 100084, China}

\collaboration{(Leading Authors)}

\author{Michael D. Albrow}
\affiliation{University of Canterbury, Department of Physics and Astronomy, Private Bag 4800, Christchurch 8020, New Zealand}

\author{Sun-Ju Chung}
\affiliation{Korea Astronomy and Space Science Institute, Daejon 34055, Republic of Korea}
\affiliation{University of Science and Technology, Korea, (UST), 217 Gajeong-ro Yuseong-gu, Daejeon 34113, Republic of Korea}

\author{Cheongho Han}
\affiliation{Department of Physics, Chungbuk National University, Cheongju 28644, Republic of Korea}

\author{Youn Kil Jung}
\affiliation{Korea Astronomy and Space Science Institute, Daejon 34055, Republic of Korea}

\author{Yoon-Hyun Ryu}
\affiliation{Korea Astronomy and Space Science Institute, Daejon 34055, Republic of Korea}

\author{In-Gu Shin}
\affiliation{Korea Astronomy and Space Science Institute, Daejon 34055, Republic of Korea}

\author{Yossi Shvartzvald}
\affiliation{Department of Particle Physics and Astrophysics, Weizmann Institute of Science, Rehovot 76100, Israel}

\author{Sang-Mok Cha}
\affiliation{Korea Astronomy and Space Science Institute, Daejon 34055, Republic of Korea}
\affiliation{School of Space Research, Kyung Hee University, Yongin, Kyeonggi 17104, Republic of Korea} 

\author{Dong-Jin Kim}
\affiliation{Korea Astronomy and Space Science Institute, Daejon 34055, Republic of Korea}

\author{Hyoun-Woo Kim}
\affiliation{Korea Astronomy and Space Science Institute, Daejon 34055, Republic of Korea}

\author{Seung-Lee Kim}
\affiliation{Korea Astronomy and Space Science Institute, Daejon 34055, Republic of Korea}
\affiliation{University of Science and Technology, Korea, (UST), 217 Gajeong-ro Yuseong-gu, Daejeon 34113, Republic of Korea}

\author{Chung-Uk Lee}
\affiliation{Korea Astronomy and Space Science Institute, Daejon 34055, Republic of Korea}

\author{Dong-Joo Lee}
\affiliation{Korea Astronomy and Space Science Institute, Daejon 34055, Republic of Korea}

\author{Yongseok Lee}
\affiliation{Korea Astronomy and Space Science Institute, Daejon 34055, Republic of Korea}
\affiliation{School of Space Research, Kyung Hee University, Yongin, Kyeonggi 17104, Republic of Korea}

\author{Byeong-Gon Park}
\affiliation{Korea Astronomy and Space Science Institute, Daejon 34055, Republic of Korea}
\affiliation{University of Science and Technology, Korea, (UST), 217 Gajeong-ro Yuseong-gu, Daejeon 34113, Republic of Korea}

\author{Richard W. Pogge}
\affiliation{Department of Astronomy, Ohio State University, 140 W. 18th Ave., Columbus, OH  43210, USA}

\collaboration{(The KMTNet Collaboration)}

\author[0000-0001-7016-1692]{Przemek Mr\'{o}z}
\affiliation{Astronomical Observatory, University of Warsaw, Al. Ujazdowskie 4, 00-478 Warszawa, Poland}
\affiliation{Division of Physics, Mathematics, and Astronomy, California Institute of Technology, Pasadena, CA 91125, USA}

\author[0000-0002-2335-1730]{Jan Skowron}
\affiliation{Astronomical Observatory, University of Warsaw, Al. Ujazdowskie 4, 00-478 Warszawa, Poland}

\author[0000-0002-9245-6368]{Radoslaw Poleski}
\affiliation{Astronomical Observatory, University of Warsaw, Al. Ujazdowskie 4, 00-478 Warszawa, Poland}

\author[0000-0002-0548-8995]{Micha{\l}~K. Szyma\'{n}ski}
\affiliation{Astronomical Observatory, University of Warsaw, Al. Ujazdowskie 4, 00-478 Warszawa, Poland}

\author[0000-0002-7777-0842]{Igor Soszy\'{n}ski}
\affiliation{Astronomical Observatory, University of Warsaw, Al. Ujazdowskie 4, 00-478 Warszawa, Poland}

\author[0000-0002-2339-5899]{Pawe{\l} Pietrukowicz}
\affiliation{Astronomical Observatory, University of Warsaw, Al. Ujazdowskie 4, 00-478 Warszawa, Poland}

\author[0000-0003-4084-880X]{Szymon Koz{\l}owski}
\affiliation{Astronomical Observatory, University of Warsaw, Al. Ujazdowskie 4, 00-478 Warszawa, Poland}

\author[0000-0001-6364-408X]{Krzysztof Ulaczyk}
\affiliation{Department of Physics, University of Warwick, Gibbet Hill Road, Coventry, CV4~7AL,~UK}

\author[0000-0002-9326-9329]{Krzysztof A. Rybicki}
\affiliation{Astronomical Observatory, University of Warsaw, Al. Ujazdowskie 4, 00-478 Warszawa, Poland}

\author[0000-0002-6212-7221]{Patryk Iwanek}
\affiliation{Astronomical Observatory, University of Warsaw, Al. Ujazdowskie 4, 00-478 Warszawa, Poland}

\author[0000-0002-3051-274X]{Marcin Wrona}
\affiliation{Astronomical Observatory, University of Warsaw, Al. Ujazdowskie 4, 00-478 Warszawa, Poland}

\author[0000-0002-1650-1518]{Mariusz Gromadzki}
\affiliation{Astronomical Observatory, University of Warsaw, Al. Ujazdowskie 4, 00-478 Warszawa, Poland}

\collaboration{(The OGLE Collaboration)}

\author{Ian A. Bond}
\affiliation{Institute of Natural and Mathematical Sciences, Massey University, Auckland 0745, New Zealand}

\author{Fumio Abe}
\affiliation{Institute for Space-Earth Environmental Research, Nagoya University, Nagoya 464-8601, Japan}

\author{Richard Barry}
\affiliation{Code 667, NASA Goddard Space Flight Center, Greenbelt, MD 20771, USA}

\author{David P. Bennett}
\affiliation{Code 667, NASA Goddard Space Flight Center, Greenbelt, MD 20771, USA}
\affiliation{Department of Astronomy, University of Maryland, College Park, MD 20742, USA}

\author{Aparna Bhattacharya}
\affiliation{Code 667, NASA Goddard Space Flight Center, Greenbelt, MD 20771, USA}
\affiliation{Department of Astronomy, University of Maryland, College Park, MD 20742, USA}

\author{Martin Donachie}
\affiliation{Department of Physics, University of Auckland, Private Bag 92019, Auckland, New Zealand}

\author{Hirosane Fujii}
\affiliation{Department of Earth and Space Science, Graduate School of Science, Osaka University, Toyonaka, Osaka 560-0043, Japan}

\author{Akihiko Fukui}
\affiliation{Department of Earth and Planetary Science, Graduate School of Science, The University of Tokyo, 7-3-1 Hongo, Bunkyo-ku, Tokyo 113-0033, Japan}
\affiliation{Instituto de Astrof\'isica de Canarias, V\'ia L\'actea s/n, E-38205 La Laguna, Tenerife, Spain}

\author{Yuki Hirao}
\affiliation{Department of Earth and Space Science, Graduate School of Science, Osaka University, Toyonaka, Osaka 560-0043, Japan}

\author{Yoshitaka Itow}
\affiliation{Institute for Space-Earth Environmental Research, Nagoya University, Nagoya 464-8601, Japan}

\author{Rintaro Kirikawa}
\affiliation{Department of Earth and Space Science, Graduate School of Science, Osaka University, Toyonaka, Osaka 560-0043, Japan}

\author{Iona Kondo}
\affiliation{Department of Earth and Space Science, Graduate School of Science, Osaka University, Toyonaka, Osaka 560-0043, Japan}

\author{Naoki Koshimoto}
\affiliation{Department of Astronomy, Graduate School of Science, The University of Tokyo, 7-3-1 Hongo, Bunkyo-ku, Tokyo 113-0033, Japan}
\affiliation{National Astronomical Observatory of Japan, 2-21-1 Osawa, Mitaka, Tokyo 181-8588, Japan}

\author{Man Cheung Alex Li}
\affiliation{Department of Physics, University of Auckland, Private Bag 92019, Auckland, New Zealand}

\author{Yutaka Matsubara}
\affiliation{Institute for Space-Earth Environmental Research, Nagoya University, Nagoya 464-8601, Japan}

\author{Yasushi Muraki}
\affiliation{Institute for Space-Earth Environmental Research, Nagoya University, Nagoya 464-8601, Japan}

\author{Shota Miyazaki}
\affiliation{Department of Earth and Space Science, Graduate School of Science, Osaka University, Toyonaka, Osaka 560-0043, Japan}

\author{Greg Olmschenk}
\affiliation{Code 667, NASA Goddard Space Flight Center, Greenbelt, MD 20771, USA}

\author{Cl\'ement Ranc}
\affiliation{Code 667, NASA Goddard Space Flight Center, Greenbelt, MD 20771, USA}

\author{Nicholas J. Rattenbury}
\affiliation{Department of Physics, University of Auckland, Private Bag 92019, Auckland, New Zealand}

\author{Yuki Satoh}
\affiliation{Department of Earth and Space Science, Graduate School of Science, Osaka University, Toyonaka, Osaka 560-0043, Japan}

\author{Hikaru Shoji}
\affiliation{Department of Earth and Space Science, Graduate School of Science, Osaka University, Toyonaka, Osaka 560-0043, Japan}

\author{Stela Ishitani Silva}
\affiliation{Department of Physics, The Catholic University of America, Washington, DC 20064, USA}
\affiliation{Code 667, NASA Goddard Space Flight Center, Greenbelt, MD 20771, USA}

\author{Daisuke Suzuki}
\affiliation{Institute of Space and Astronautical Science, Japan Aerospace Exploration Agency, 3-1-1 Yoshinodai, Chuo, Sagamihara, Kanagawa, 252-5210, Japan}

\author{Yuzuru Tanaka}
\affiliation{Department of Earth and Space Science, Graduate School of Science, Osaka University, Toyonaka, Osaka 560-0043, Japan}

\author{Paul J. Tristram}
\affiliation{University of Canterbury Mt.\ John Observatory, P.O. Box 56, Lake Tekapo 8770, New Zealand}

\author{Tsubasa Yamawaki}
\affiliation{Department of Earth and Space Science, Graduate School of Science, Osaka University, Toyonaka, Osaka 560-0043, Japan}

\author{Atsunori Yonehara}
\affiliation{Department of Physics, Faculty of Science, Kyoto Sangyo University, 603-8555 Kyoto, Japan}

\collaboration{(The MOA Collaboration)}

\author{Charles A. Beichman}
\affiliation{IPAC, Mail Code 100-22, Caltech, 1200 E. California Blvd., Pasadena, CA 91125, USA}

\author{Geoffery Bryden}
\affiliation{Jet Propulsion Laboratory, California Institute of Technology, 4800 Oak Grove Drive, Pasadena, CA 91109, USA}

\author{Sebastiano Calchi Novati}
\affiliation{IPAC, Mail Code 100-22, Caltech, 1200 E. California Blvd., Pasadena, CA 91125, USA}

\author{Sean Carey}
\affiliation{IPAC, Mail Code 100-22, Caltech, 1200 E. California Blvd., Pasadena, CA 91125, USA}

\author[0000-0003-0395-9869]{B. Scott Gaudi}
\affiliation{Department of Astronomy, Ohio State University, 140 W. 18th Ave., Columbus, OH  43210, USA}

\author{Calen B. Henderson}
\affiliation{IPAC, Mail Code 100-22, Caltech, 1200 E. California Blvd., Pasadena, CA 91125, USA}

\author{Samson Johnson}
\affiliation{Department of Astronomy, Ohio State University, 140 W. 18th Ave., Columbus, OH  43210, USA}

\collaboration{(The \emph{Spitzer} Team)}

%% file: abstract.tex
\begin{abstract}
In order to exhume the buried signatures of ``missing planetary caustics'' in the KMTNet data, we conducted a systematic anomaly search to the residuals from point-source point-lens fits, based on a modified version of the KMTNet EventFinder algorithm. This search reveals the lowest mass-ratio planetary caustic to date in the microlensing event OGLE-2019-BLG-1053, for which the planetary signal had not been noticed before. The planetary system has a planet-host mass ratio of $q = (1.25 \pm 0.13) \times 10^{-5}$. A Bayesian analysis yields estimates of the mass of the host star, $M_{\rm host} = 0.61_{-0.24}^{+0.29}~M_\odot$, the mass of its planet, $M_{\rm planet} = 2.48_{-0.98}^{+1.19}~M_{\earth}$, the projected planet-host separation, $a_\perp = 3.4_{-0.5}^{+0.5}$~au, and the lens distance of $D_{\rm L} = 6.8_{-0.9}^{+0.6}$~kpc. The discovery of this very low mass-ratio planet illustrates the utility of our method and opens a new window for a large and homogeneous sample to study the microlensing planet-host mass-ratio function down to $q \sim 10^{-5}$. 

\end{abstract}

%% file: intro.tex
The structure of the caustics plays a central role in the phenomenology of planetary microlensing light curves and thus the detectability of microlensing planets. A source must transit or come close to a caustic to create a detectable signal \citep{Shude1991, Andy1992, Gaudi2012}. Planetary companions to microlensing hosts induce three classes of caustic structures: central, planetary and resonant caustics. For $s > s_w$ or $s < s_c$, where $s$ is the planet-host separation in units of the Einstein radius $\thetae$, $s_w \simeq 1 + 3q^{1/3}/2$, $s_c \simeq 1 - 3q^{1/3}/4$ and $q$ is the planet-host mass ratio \citep{Dominik1999}, the caustic structure consists of a small quadrilateral caustic near the host (central caustic) and one quadrilateral (for $s > s_w$) or two triangular (for $s < s_c$) caustics separated from the host position by $|s - s^{-1}|\thetae$ (planetary caustics). For $s_c < s < s_w$, the central and planetary caustics merge together and form a 6-sided ``resonant'' caustic near the host. \cite{OB190960} showed that ``near-resonant'' caustics, which have boundaries ($-3\log s_c, 1.8\log s_w$), are as sensitive as resonant caustics due to their long magnification ridges (or troughs) extending from the central caustic and the planetary caustics. For a clear definition, we refer to caustics out of the near-resonant range as ``pure-planetary'' caustics. 

Although resonant and near-resonant caustics occupy a relatively narrow range of $s$, more than 80 of microlensing planets were detected via these two classes of caustics, while only 25 microlensing planets were discovered by ``pure-planetary'' caustics. See the $\log q$ vs.\ $\log s$ plot for the 114 published microlensing planets in Figure \ref{qs}. Besides the high intrinsic sensitivity of resonant and near-resonant caustics, detection bias plays an important role. For many years (beginning with the second microlens planet, OGLE-2005-BLG-071Lb, \citealt{OB050071}), $\gtrsim 2/3$ of microlensing planets (see Figure 10 of \citealt{OB160596}) were discovered based on the two-step approach advocated by \cite{Andy1992}. In the first step, because the typical Einstein timescale $\te$ for microlensing events is about $20$~days \citep{Mroz2017a}, a wide-area survey with a cadence of $\Gamma \sim 1~{\rm day}^{-1}$ is sufficient to find microlensing events. In the second step, individual events found in the first step would be monitored by high-cadence follow-up observations from a broadly distributed network, in order to characterize the planetary signal \citep{PLANET,mufun,RoboNet,MINDSTEp}. Due to the scarcity of telescope resources and the fact that the peak of an event can usually be predicted in advance, follow-up observations were most successful when they focused on the peak of high-magnification events, for which the source trajectory goes close to the host. Because of the large caustic size and the long magnification ridges near the host, sources of high-magnification events frequently intersect resonant and near-resonant caustics, and this explains the high frequency of microlensing planets detected through this channel. In the non-resonant case, in which the central and planetary caustics are well detached, the size of the central caustic scales as $\propto s^2$ for $s < 1$ and $\propto s^{-2}$ for $s > 1$ \citep{Chung2005}, which requires dense coverage over the peak of very-high-magnification (and therefore rare) events to capture the planetary signal, and thus only six such planets have been detected via this channel\footnote{The six planets are OGLE-2006-BLG-109Lc \citep{OB06109, OB06109_Dave}, OGLE-2007-BLG-349Lb \citep{OB07349}, MOA-2007-BLG-400Lb \citep{MB07400,MB07400_AO}, MOA-2011-BLG-293Lb \citep{MB11293}, OGLE-2012-BLG-0563Lb \citep{OB120563} and OGLE-2013-BLG-0911Lb \citep{OB130911}}.

For the broad range of pure-planetary caustics, random source trajectories intersect the planetary caustic(s) much more often than the central caustic. For $s > 1$, the ratio between the size of planetary/central caustics is $\sim q^{-1/2}$ \citep{Han2006}, and hence the planetary caustic is about 100 times larger than the central caustic for the common $q \sim 10^{-4}$ planets \citep[e.g.,][]{OB05390}. For $s < 1$, the ratio is $\sim 0.3q^{-1/2}s$ \citep{Han2006}, and hence the two planetary caustics are an order of 10 times larger than the central caustic for $q \sim 10^{-4}$. Thus, the planetary caustic can play an important role in microlensing planet detections, especially for low mass-ratio planets, provided that high-cadence observations for the whole light curves can be conducted. The Microlensing Observations in Astrophysics (MOA, one 1.8 m telescope equipped with a 2.4 ${\rm deg}^2$ camera at New Zealand, \citealt{MOA2016}) and the Optical Gravitational Lensing Experiment (OGLE, one 1.3 m telescope equipped with a 1.4 ${\rm deg}^2$ camera at Chile, \citealt{OGLEIV}) were the first to cover wide areas with high cadences of $\Gamma = 1-4~{\rm hr}^{-1}$, which enables the detection of both microlensing events and microlensing planets without the need for follow-up observations for many events. The detection rate of pure-planetary caustics rapidly increased with the upgrades of the OGLE and MOA experiments, including the lowest mass-ratio planet prior to 2018, OGLE-2013-BLG-0341Lb with $q = (4.43 \pm 0.029) \times 10^{-5}$ \citep{OB130341}.

The new-generation microlensing survey, the Korea Microlensing Telescope Network (KMTNet, \citealt{KMT2016}), consists of three 1.6 m telescopes equipped with $4~{\rm deg}^2$ cameras at the Cerro Tololo Inter-American Observatory (CTIO) in Chile (KMTC), the South African Astronomical Observatory (SAAO) in South Africa (KMTS), and the Siding Spring Observatory (SSO) in Australia (KMTA). Beginning in 2016, KMTNet conducted near-continuous observations for a total area of about $100~ {\rm deg}^2$ toward the Galactic bulge, with about $12~{\rm deg}^2$ at a high cadence of $\Gamma \sim 4~{\rm hr}^{-1}$, and about $28~{\rm deg}^2$ at a high cadence of $\Gamma \sim 1~{\rm hr}^{-1}$. The enhanced observational cadence of the KMTNet survey resulted in the great increase of the planet detection rate, and the microlensing planets detected with the KMTNet data comprise about half of all published planets despite of its short period of operation (see the red points in Figure \ref{qs}).

\cite{Zhu2014ApJ} simulated a KMTNet-like survey and found that more than half of KMT $q < 10^{-4}$ planets should be detected via the channel of pure-planetary caustics (see their Figure 4). In contrast to this prediction, the KMT planets detected through the channel of pure-planetary caustics 
comprise a minor fraction of all planet sample. Here we define this discrepancy as ``missing planetary caustics'' problem. Among the 14 KMT $q < 10^{-3}$ planets, only two were detected by pure-planetary caustics, OGLE-2018-BLG-0596Lb \citep{OB180596} with $q \sim 2 \times 10^{-4}$ and OGLE-2017-BLG-0173Lb with $q \sim (2~{\rm or}~6) \times 10^{-5}$ \citep{OB170173}. Among the 29 $q < 10^{-3}$ planets without KMT data, eight have pure-planetary caustics, while follow-up observations on high-magnification events played an important role in the detections of resonant and near-resonant caustics \citep[e.g.,][]{OB05169}\footnote{The two lowest mass-ratio KMT planets, OGLE-2019-BLG-0960Lb and KMT-2020-BLG-0414Lb, were detected by joint observations of surveys and follow-up teams. For OGLE-2019-BLG-0960Lb, although the planetary signal was first recognized by the follow-up data, the KMT-only data were sufficient to discover the planet (see Section 6.1 of \citealt{OB190960}). For KMT-2020-BLG-0414Lb, KMTC and KMTS were closed due to Covid-19. However, because the planetary signal lasted for about five days, KMT-only would have been able to detect the planet if KMTC and KMTS had been open \cite{KB200414}.}

The ``missing planetary caustics'' in the KMT $q < 10^{-3}$ planet sample could be due to the way that we search for planetary signals. Although KMTNet + OGLE + MOA conduct high-cadence observations over the whole microlensing season, the systematic search for planetary signals has not been extended to the light curves of whole events. For most events, modelers only search for anomalies by a visual inspection of the light curve, with their main attention devoted to the peak. For high-magnification events which are intrinsically more sensitive to planets, modelers may carefully check the observed data of the peak and the residuals from a point-source point-lens (PSPL, \citealt{Paczynski1986}) fit \citep[e.g.,][]{KB190842,KB181025}, and even trigger tender-loving care (TLC) re-reductions \citep[e.g.,][]{KB191953}. However, the signals of $q < 10^{-3}$ planetary caustics generally occur on the wings of light curves, with low amplitudes and large photometric uncertainties, and thus could have been missed due to human bias (i.e., focus on the near-peak region).

In order to find the ``missing planetary caustics'', we conducted a systematic anomaly search to the whole annual light curve. We applied a modified version of the KMT EventFinder algorithm \citep{KMTeventfinder} to the residuals from PSPL fits and found the lowest mass-ratio planetary caustic to date in the event \event, with $q = (1.25 \pm 0.13) \times 10^{-5}$. 

The paper is structured as follows. In Section \ref{anomaly}, we describe the basic algorithm and procedures for the anomaly search. We then introduce the observations, the light-curve analysis and the physical parameters of \event\ in Sections \ref{obser}, \ref{model} and \ref{lens}, respectively. Finally, we discuss the implications of our work in Section \ref{dis}.

%% file: anomaly.tex
\subsection{Basic Algorithm}

Normally, an anomaly in a microlensing curve refers to a deviation from a PSPL model, which could be of astrophysical origin such as an additional lens (2L1S, \citealt{Shude1991}), an additional source (1L2S, \citealt{Griest1992}) or finite-source effects \citep{1994ApJ...421L..75G,Shude1994,Nemiroff1994}, or caused by artifacts. For most microlensing planetary events, the planet-mass companion only induces several-hour to several-day deviations to a PSPL model, and the residuals from a PSPL model fit a zero-flux flat curve with short-lived deviations in some places. Thus, our basic idea is to search for such short deviations from the residuals to a PSPL model.

\cite{Wise} first applied an anomaly search algorithm to real complete light curves (OGLE + MOA + Wise). They calculated local $\chi^2$ for every 40 points and select an anomaly if a local $\chi^2$ exceeds a threshold. However, this algorithm does not consider the correlations in the residuals induced by by real anomalies, and it results in many false positives due to the systematics of KMT end-of-year-pipeline light curves. Thus, we apply the KMT EventFinder algorithm \citep{KMTeventfinder} for the anomaly search. The KMT EventFinder adopts a \cite{Gould2D} 2-dimensional (2D) grid of $(t_0, \teff)$ to search for microlensing events in the KMT end-of-year-pipeline light curves, where $\teff = u_0\te$ is the effective timescale, $t_0$ is the time of the maximum magnification, $u_0$ is the impact parameter in units of the angular Einstein radius $\thetae$, and $\te$ is the Einstein radius crossing time \citep{Paczynski1986}. It uses two approaches to fit the observed flux, $F(t)$,
\begin{equation}
    F(t) = f_1 A_j[Q(t;t_0,\teff)] + f_0; \qquad Q(t;t_0,\teff) \equiv 1 + \biggl(\frac{t - t_0}{\teff}\biggl)^2; \qquad (j = 0, 1)
\end{equation}
where
\begin{equation}
    A_{j=0}(Q) = Q^{-1/2}; \qquad A_{j=1}(Q) = \frac{Q + 2}{\sqrt{Q(Q + 4)}} = [1 - (Q/2 + 1)^{-2}]^{-1/2},
\end{equation}
and ($f_1, f_0$) are two flux parameters, which are evaluated by a linear fit.

In reality, the planetary deviations are not simply symmetric single ``bumps'' except for events that consist of two isolated PSPL curves that are respectively caused by the host and a wide-orbit planet \citep[e.g.,][]{OB160263}, so our search model cannot fit the deviations perfectly. However, the main purpose of the 2D grid search is to locate the signal and roughly estimate its significance. For a signal that passes the EventFinder threshold, the KMT EventFinder pipeline further fits it with a PSPL model and evaluates it with a second threshold \citep{KB192073}. Given an acceptable level of effort to carry out a manual review with low-threshold candidates (see Section \ref{manual} and \ref{future}), it is unnecessary to design models that perfectly fit the light curve, which would actually be very difficult due to the diversity of deviations. In addition, the deviations contain not only ``bumps'', which are the targets of the EventFinder, but also ``dips'' \citep[e.g.,][]{OB130341} and ``U shapes'', which are caused by caustic crossings \citep[e.g.,][]{OB03235}. Nevertheless, ``dips'' can be regarded as the inverse of ``bumps'' and be fitted by a negative $f_1$, while each peak of ``U shapes'' or even the whole ``U shapes'' can be regarded as a bump, as shown in Figure 11 of \cite{KMTeventfinder}.

\subsection{Data Handling}

KMTNet made end-of-year-pipeline light curves public for the 2015--2019 seasons\footnote{http://kmtnet.kasi.re.kr/~ulens/}. We adopt the events from the 2019 season, because its light-curve files contain seeing and sky background information. This auxiliary information provides a systematic way to exclude most of the bad points which frequently generate fake signals. Based on an investigation of bad points, we exclude data points that have a sky background brighter than $5000$ ADU/pixel\footnote{For the KMTNet cameras, the gain is 1.0 photo-electrons per analog-to-digital unit (ADU)} or a seeing FWHM larger than 7 pixels ($0.4^{\prime\prime}$ per pixel) for the KMTA and KMTS data and 6.5 pixels for the KMTC data. We also exclude KMTS data between ${\rm HJD}^{\prime}$ = 8640 -- 8670 (${\rm HJD}^{\prime} = {\rm HJD} - 2450000$) on CCD N chip, which have anomalous fluxes due to a failing electrical connection in that chip.

In general, the errors from photometric measurements for each data set $i$ were renormalized using the formula $\sigma_i^{\prime} = k_i\sqrt{\sigma_i^2 + e^2_{i,{\rm min}}}$, where $\sigma_i$ and $\sigma_i^{\prime}$ are the original error bars from the photometry pipelines and the renormalized error bars in magnitudes, and $k_i$ and $e_{i,{\rm min}}$ are rescaling factors. The rescaling factors are often determined using the method of \cite{MB11293}, which enables $\chi^2/{\rm dof}$ for each data set to become unity. However, this procedure is not feasible for our search. For the PSPL fits, the error bars were overestimated, because some outliers have not been excluded by the seeing and sky background thresholds, and the data cannot fit a PSPL model if an event includes an anomaly. For the anomaly search to the residuals, because our search model cannot fit the deviations perfectly, it is unreasonable to require $\chi^2/{\rm dof} = 1$. Thus, we simply adopt $k = 1.5$ and $e_{\rm min} = 0$ for each data set, after an investigation of the rescaling factors of error bars for a subset of PSPL events.

Finally, the pipeline data, which are in the magnitude units, are converted to the flux unit using the same $(I = 28)$ zero point that was used by the KMT end-of-year pipeline.


\subsection{Event Selection}

We adopt the $\Icat < 19.0$ events as our first sample (1216 in total), where $\Icat$ is the star-catalog magnitude entry in the KMT database. For regions covered by the OGLE-III catalog \citep{OGLEIII}, we adopt the $\Icat$ value from the OGLE-III catalog. For most regions that are not covered by OGLE-III, $\Icat$ is taken as the $i^\prime$ magnitude from the catalog of \cite{DECam} derived from DECam data. For the small regions not covered by either catalog, $\Icat$ is derived from \texttt{DoPHOT} \citep{dophot} reductions of the KMT templates. For each event, the KMT end-of-year-pipeline adopts $\Icat$ as the baseline magnitude of light curves, which includes both the source flux and the blended light. There are two reasons for this brightness threshold. First, the main purpose of the current search is to develop and test the method and programming, which requires repeated computation and manual review. To ease the burden, it is necessary to select a small but sensitive sample. Second, because the signals of planetary caustics often occur on the wings of the light curves and the $I \geq 19.0$ data have large photometric uncertainties, it is difficult (but not impossible, e.g., \citealt{OB151771}) to find planetary signals from the $I \geq 19.0$ data. A more comprehensive approach may be to adopt all of the $I < 19.0$ data, rather than selecting only $I_{\rm cat} < 19.0$ events, but the current sample is sufficient for the main purpose of our search. We will discuss further improvements to our search in Section \ref{future}. 

We fit the 1216 $I_{\rm cat} < 19.0$ events with the PSPL model by a downhill\footnote{We use a function based on the Nelder-Mead simplex algorithm from the SciPy package. See \url{https://docs.scipy.org/doc/scipy/reference/generated/scipy.optimize.fmin.html\#scipy.optimize.fmin}} approach using ($t_0, u_0, \te$) from the KMT website as the initial parameters. We then manually review the PSPL model plots and find that 219 events have either an obvious variable source, too low signal-to-noise ratios of the microlensing effects, very noisy photometry for all of the data sets, or are of non-microlensing origins (e.g., cataclysmic variables). We remove these events. For the remaining 997 events, we photometrically align the PSPL residuals of each data set to the KMTC or KMTS residuals using the two flux parameters, ($f_1, f_0$), from the PSPL fits.

\subsection{Detailed Search}
The set of $t_{{\rm eff},k}$ are a geometric series,
\begin{equation}
    t_{{\rm eff},k+1} = (4/3)t_{{\rm eff},k},
\end{equation}
with the shortest effective timescale $t_{{\rm eff},1} = 0.30$ days and the longest effective timescale $t_{{\rm eff},13} = 0.30\times(4/3)^{12} = 9.47$ days. Here $t_{{\rm eff},1} = 0.30$ is adopted from the current lower limit of $\teff$ of the KMT EventFinder pipeline \citep{KB192073}. While $\teff \gtrsim 5$ days is definitely too long for planetary signals, we consider that some short-timescale events could be caused by a wide-orbit planet \citep[e.g.,][]{OB161227}, so the series of long $\teff$ are designed for the weak signals of a possible host star. The step size of $t_0$ is $\delta_{t_0} = (1/6)\teff$, and the grids begin at $\delta_{t_0}$ before the first epoch of the 2019 season and end at $\delta_{t_0}$ after the last epoch. We restrict the search at each grid point ($t_0, \teff$) to data within $t_0 \pm 3~\teff$ and require that this interval contains at least five data points and at least three successive points $\geq 2\sigma$ away from the zero-residual curve. 

Finally, each grid point is evaluated by two parameters, $\Delta\chi^2_{\rm zero}$ and $\Delta\chi^2_{\rm flat}$,
\begin{equation}
    \Delta\chi^2_{\rm zero} = \chi^2_{\rm zero} - \chi^2_{\rm signal}; \qquad \Delta\chi^2_{\rm flat} = \chi^2_{\rm flat} - \chi^2_{\rm signal},
\end{equation}
where $\chi^2_{\rm zero}$, $\chi^2_{\rm flat}$ and $\chi^2_{\rm signal}$ are the $\chi^2$ to the zero-flux curve, the mean-flux curve, and the search model, respectively, $\Delta\chi^2_{\rm zero}$ determines the significance of the signal, and $\Delta\chi^2_{\rm flat}$ characterizes the steepness of the residual flux. For most signals, such as clear ``bumps'' or ``dips'', both $\Delta\chi^2_{\rm zero}$ and $\Delta\chi^2_{\rm flat}$ are significant. However, for some long-$\teff$ signals that are caused by long-term variability or systematics, $\Delta\chi^2_{\rm flat} \sim 0$. After reviewing some recognized signals with different $\Delta\chi^2_{\rm zero}$ and $\Delta\chi^2_{\rm flat}$, we decide to select if (1) $\Delta\chi^2_{\rm zero} > 120$; or (2) $\Delta\chi^2_{\rm zero} > 75$ and $\Delta\chi^2_{\rm flat} >35$. After reviewing some recognized signals, we find that in average $\chi^2_{\rm 1L1S} - \chi^2_{\rm 2L1S} \sim 1.6$ $\Delta\chi^2_{\rm zero}$. Taking into account that it could require $\chi^2_{\rm 1L1S} - \chi^2_{\rm 2L1S} \sim 120$ to ensure a real detection and the efforts required for manual review, we decide to select if (1) $\Delta\chi^2_{\rm zero} > 120$; or (2) $\Delta\chi^2_{\rm zero} > 75$ and $\Delta\chi^2_{\rm flat} >35$. Two signals (A, B) from the same event are judged to be the same signal provided that $|t_{0,{\rm A}} -t_{0,{\rm B}}| < t_{\rm eff, A} + t_{\rm eff, B}$. As a result, the anomaly search yielded 6320 candidate signals from 422 events.  

\subsection{Manual Review and Results}\label{manual}

Each candidate is shown to the operator in a four-panel display together with some auxiliary information. The display shows the light curves and residuals for the signal and for the data of the whole season. See Figure \ref{review} for an example. For candidates that are assessed as plausibly real (i.e., not an artifact), the operator first checks whether the event was independently found by OGLE and/or MOA, and if so whether their on-line light curves have data points during the anomaly. If they do, and if these data points are inconsistent with the KMT-based anomaly, the candidate is rejected. For example, for KMT-2019-BLG-0607/OGLE-2019-BLG-0667, the KMTC data shows a $\sim 0.3$-day bump on the peak, but the OGLE data do not show this bump. If no such external check is possible, then the anomaly is investigated by a variety of techniques at the image level before proceeding to the next step.  For example, for KMT-2019-BLG-2418, a long, low-amplitude bump was found about 120 days before the $\te \sim 4$-day short event that had previously been selected as a microlensing event. The bump appeared in all three KMT data sets, and so could have represented a ``host'' to the short-event ``planet''. Neither OGLE nor MOA had found a counterpart to this event. However, investigation of the images showed that the bump was due to flux from a nearby variable, so the candidate was rejected.

As a result, the operator (W. Zang) identified 24 candidates that could be planetary events and 59 candidates that should be other types of anomaly (e.g., binary-star events). Among the 24 candidate planets, four are known planets \citep[e.g.,][]{OB190960} and four are finite-source point-lens events \citep{KB192073}. For the remaining 16 candidates, preliminary 2L1S fits suggested that \event\ has a pure-planetary caustic induced by a very low mass-ratio planet. This triggered TLC re-reductions for the KMT data, which combined with the OGLE data on the anomaly, revealed a clear planetary signal. 



%% file: obser.tex
On 5 July 2019, \event\ was announced as a microlensing candidate event by the OGLE Early Warning System \citep{Udalski1994,Udalski2003} at equatorial coordinates $(\alpha, \delta)_{\rm J2000}$ = (18:00:39.93, $-27$:20:29.7), corresponding to Galactic coordinates $(\ell,b)=(3.06, -2.05)$. It was then independently discovered by the KMT alert-finder system \citep{KMTAF} at the position of an $I = 18.84$ catalog star and announced as a clear microlensing candidate KMT-2019-BLG-1504 on 7 July 2019.  

The OGLE observations were carried out using the 1.3 m Warsaw Telescope equipped with a 1.4 ${\rm deg}^2$ FOV mosaic CCD camera at the Las Campanas Observatory in Chile \citep{OGLEIV}. The event lies in the OGLE BLG511 field, with a cadence of $\Gamma = 1~{\rm hr}^{-1}$. The event lies in two slightly offset KMT fields, BLG03 and BLG43, with a combined cadence of $\Gamma \sim 4~{\rm hr}^{-1}$. 

For both surveys, most images were taken in the $I$ band, and a fraction of images were taken in the $V$ band for the source color measurements. In addition, This event was also observed by the \Sp\ space telescope. We discuss those observations
in Appendix \S~\ref{sec:spitzer}.

The ground-based data used in the light curve analysis were reduced using custom implementations of the difference image analysis technique \citep{Tomaney1996,Alard1998}: \cite{Wozniak2000} for the OGLE data and pySIS \citep{pysis} for the KMT data. For the KMTC03 data, we conduct pyDIA photometry\footnote{MichaelDAlbrow/pyDIA: Initial Release on Github, doi:10.5281/zenodo.268049} to measure the source color. The $I$-band magnitude of the data has been calibrated to the standard $I$-band magnitude using the OGLE-III star catalog \citep{OGLEIII}. The errors from photometric measurements for each data set were readjusted following the routine of \cite{MB11293}. The data used in the analysis, together with the corresponding data reduction method and the rescaling factors are summarized in Table \ref{data}.

%% file: model.tex
\subsection{Heuristic Analysis}\label{estimate}

Figure \ref{lc} shows the \event\ data together with the best-fit models. The light curve shows two consecutive small bumps ($t_{0,{\rm anom}} \sim 8670$) 20.5 days before the peak of an otherwise normal PSPL light curve. Such a bump is a typical signature of a planet produced by the source approach or crossing over the planetary caustic \citep{Andy1992}. The 2L1S model requires three additional parameters $(s, q, \alpha)$, where $\alpha$ is the angle of the source trajectory relative to the binary axis. We also consider finite-source effects and include the source radius normalized by the Einstein radius, $\rho = \theta_*/\thetae$.  

We first fit the PSPL model excluding the data points around the anomaly and obtain 
\begin{equation}
    (t_0, u_0, \te) = (8690.5, 0.35, 34~{\rm days}),
\end{equation}
which leads to
\begin{equation}
 \tau_{\rm anom} = \frac{t_{\rm anom} - t_0}{\te} = -0.60; \qquad u_{\rm anom} = \sqrt{u_0^2 + \tau_{\rm anom}^2} = 0.695; \qquad  |\alpha| = |\tan^{-1}\frac{u_0}{\tau_{\rm anom}}| = 0.53~(30.3^\circ).
\end{equation}
Because the planetary caustic is located at the position of $|s - s^{-1}| \sim u_{\rm anom}$, we obtain
\begin{equation}
    s \sim \frac{\sqrt{u_{\rm anom}^2 + 4} + u_{\rm anom}}{2} = 1.41~{\rm or}~s \sim \frac{\sqrt{u_{\rm anom}^2 + 4} - u_{\rm anom}}{2} = 0.71.
\end{equation}
For the remaining two 2L1S parameters, $q$ and $\rho$, a systematic search is required.  

\subsection{Numerical Analysis}
We use the advanced contour integration code \citep{Bozza2010,Bozza2018} \texttt{VBBinaryLensing}\footnote{\url{http://www.fisica.unisa.it/GravitationAstrophysics/VBBinaryLensing.htm}} to calculate the magnification of the 2L1S model. We locate the $\chi^2$ minima by conducting a grid search over the parameter plane ($\log s, \log q, \alpha$). The grid consists of 21 values equally spaced between $-0.2\leq\log s \leq0.2$, 10 values equally spaced between $0^{\circ}\leq \alpha < 360^{\circ}$, and 61 values equally spaced between $-6\leq \log q \leq0$. For each set of ($\log s, \log q, \alpha$), we fix $\log q$, $\log s$ and let the other parameters ($t_0, u_0, \te, \rho, \alpha$) vary. We find a lensing solution using the Markov chain Monte Carlo (MCMC) $\chi^2$ minimization applying the \texttt{emcee} ensemble sampler \citep{emcee}. From this, we find two distinct minima with ($\log s, \log q) \sim$ ($-0.15, -4.5$) and ($0.15, -4.9$) and label them by ``Close'' ($s < 1$) and ``Wide'' ($s > 1$) in the following analysis. We then investigate the best-fit models with all free parameters. The best-fit parameters with their $68\%$ uncertainty range from the MCMC are shown in Table \ref{parm1}, and the caustics and source trajectories are shown in Figure \ref{cau}. We note that the heuristic estimates for $(s,\alpha)$ are in good agreement with the values in Table \ref{parm1}.

We found that the Wide model provides the best fit to the observed data, and the $\chi^2$ improvement to the best-fit PSPL model is 453.6. The two consecutive small bumps are produced by the source crossing the two spikes of the quadrilateral caustic. The Close model is disfavored by $\Delta\chi^2 = 41.6$, and all of the $\chi^2$ difference come from the anomalous region. We also check whether the $\Delta \chi^2$ can be decreased by considering the microlens ground-based parallax effect \citep{Gould1992, Gould2000, Gouldpies2004}, which is caused by the orbital acceleration of Earth, and the lens orbital motion effect \citep{MB09387,OB09020}, but all the Close solutions have $\Delta\chi^2 > 40$ compared to the Wide solutions and cannot reproduce the double-bump feature. Thus, we exclude the Close model and only investigate the Wide model in the following analysis. In addition, we check the 1L2S model and find that it is disfavored by $\Delta\chi^2 > 400$. Thus, we exclude the 1L2S model, too. 

We check whether the fit to the wide model further improves by including the microlens ground-based parallax effect,
\begin{equation}
\bm{\pi}_{\rm E} = \frac{\pi_{\rm rel}}{\thetae}\frac{\bm{\mu}_{\rm rel}}{\mu_{\rm rel}}~,   
\end{equation}
where $(\pi_{\rm rel}, \bm{\mu}_{\rm rel})$ are the lens-source relative (parallax, proper motion). We parameterize the microlens parallax by $\pi_{\rm E,N}$ and $\pi_{\rm E,E}$, which are the North and East components of the microlens parallax vector. We also fit the $u_0 > 0$ and $u_0 < 0$ solutions to consider the ``ecliptic degeneracy'' \citep{Jiang2004, Poindexter2005}. The addition of parallax to the model only improves $\Delta\chi^2 \leq 1.2$ (see Table \ref{parm1}), but it provides ``1-D parallax'' constraint with $\sigma(\pi_{\rm E, \parallel}) \sim 0.05$, where $\pi_{\rm E, \parallel}$ is the component of $\bm{\pi}_{\rm E}$ that in the direction of the projected position of the Sun at $t_0$. We also consider the lens orbital motion effect and find that it is not detectable ($\Delta\chi^2 < 0.3$) and not correlated with $\bm{\pi}_{\rm E}$, so we eliminate the lens orbital motion from the fit.


%% file: lens.tex
\subsection{Color Magnitude Diagram}\label{CMD}
We estimate the intrinsic brightness and color of the source by locating the source on a color magnitude diagram (CMD) \citep{Yoo2004}. We construct a $V - I$ versus $I$ CMD using the OGLE-III catalog stars \citep{OGLEIII} within $80^{\prime\prime}$ centered on the event (see Figure \ref{cmd}). We measure the centroid of the red giant clump as $(V - I, I)_{\rm cl} = (2.45 \pm 0.01, 16.11 \pm 0.02)$ and adopt the intrinsic color and de-reddened magnitude of the red giant clump $(V - I, I)_{\rm cl,0} = (1.06, 14.35)$ from \cite{Bensby2013} and \cite{Nataf2013}. For the source color, we obtain $(V - I)_{\rm S} = 2.09 \pm 0.03$ by regression of the KMTC03 $V$ versus $I$ flux with the change of the lensing magnification and a calibration to the OGLE-III magnitudes. Using the color/surface-brightness relation for dwarfs and subgiants of \cite{Adams2018}, we obtain 
\begin{numcases}{\theta_* =}
0.762 \pm 0.053 ~\mu {\rm as}~{\rm for~the}~u_0 > 0~{\rm solution}, \\
0.759 \pm 0.053 ~\mu {\rm as}~{\rm for~the}~u_0 < 0~{\rm solution}.      
\end{numcases}

\subsection{Bayesian Analysis}\label{Baye}

For a lensing object, the total mass $M_{\rm L}$ and the lens distance $D_{\rm L}$ are related to the angular Einstein radius $\thetae$ and the microlens parallax $\pie$ by \citep{Gould1992, Gould2000}
\begin{equation}\label{eq:mass}
    M_{\rm L} = \frac{\thetae}{{\kappa}\pie};\qquad D_{\rm L} = \frac{\mathrm{au}}{\pie\thetae + \pi_{\rm S}},
\end{equation}
where $\kappa \equiv 4G/(c^2\mathrm{au}) = 8.144$ mas$/M_{\odot}$, $\pi_{\rm S} = \mathrm{au}/D_{\rm S}$ is the source parallax, and $D_{\rm S}$ is the source distance. Using the measurements of $\rho$ from the light-curve analysis and $\theta_*$ from the CMD analysis, we obtain the angular Einstein radius
\begin{numcases}{\thetae = \frac{\theta_*}{\rho} = }
0.366 \pm 0.039 ~\text{mas}~{\rm for~the}~u_0 > 0~{\rm solution}, \\
0.367 \pm 0.039 ~\text{mas}~{\rm for~the}~u_0 < 0~{\rm solution}.      
\end{numcases}
Combined with the measurement $\te \sim 34$~days, these values imply a lens-source relative proper motion $\mu_{\rm rel} \sim 4~{\rm mas\,yr^{-1}}$. However, the observed data only give a weak constraint on the microlens parallax. We therefore conduct a Bayesian analysis based on a Galactic model to estimate the physical parameters of the planetary system.

The Galactic model mainly consists of three aspects: the mass function of the lens, the stellar number density profile and the source and lens velocity distributions. For the lens mass function, we begin with the initial mass function (IMF) of \cite{Kroupa2001} for both the disk and the bulge. To approximate the impact of the age and vertical dispersion as a function of age of the disk population, we impose a cut off of $1.3~M_{\odot}$ \citep{Zhu2017spitzer}. Taking account of the age distribution of microlensed dwarfs and subgiants of Figure 13 of \cite{Bensby2017}, we impose a cut off of $1.1~M_{\odot}$ for the bulge. For the bulge and disk stellar number density, we choose the models used by \cite{Zhu2017spitzer} and \cite{MB11262}, respectively. For the disk velocity distribution, we assume the disk lenses follow a rotation of $240~{\rm km~s}^{-1}$ \citep{Reid2014} with the velocity dispersion of \cite{KB180748}. For the bulge dynamical distributions, we adopt the {\it Gaia} proper motion of red giant stars within $5'$ \citep{Gaia2016AA,Gaia2018AA} and obtain 
\begin{equation}
\langle\bm{\mu}_{\rm bulge}(\ell, b)\rangle = (-5.65, -0.09) \pm (0.15, 0.11)~\text{mas yr}^{-1},
\end{equation}
\begin{equation}
\sigma(\bm{\mu}_{\rm bulge}) = (3.15, 2.54) \pm (0.17, 0.13) ~\text{mas yr}^{-1}.    
\end{equation}

We create a sample of $10^8$ simulated events from the Galactic model. For each simulated event $i$ of solution $k$, we weight it by
\begin{equation}\label{eq:weight}
    \omega_{{\rm Gal},i,k} = \Gamma_{i,k} \mathcal{L}_{i,k}(\te) \mathcal{L}_{i,k}(\thetae) \mathcal{L}_{i,k}(\bm{\pi}_{\rm E}) ,
\end{equation}
where $\Gamma_{i,k}\varpropto\theta_{{\rm E},i,k}\times\mu_{{\rm rel},i,k}$ is the microlensing event rate, $\mathcal{L}_{i,k}(\te)$, $\mathcal{L}_{i,k}(\thetae)$ and $\mathcal{L}_{i,k}(\bm{\pi}_{\rm E})$ are the likelihood of its inferred parameters $(\te, \thetae, \bm{\pi}_{\rm E})_{i,k}$ given the error distributions of these quantities derived from the MCMC for that solution
\begin{equation}
    \mathcal{L}_{i,k}(\te) = \frac{{\rm exp}[-(t_{{\rm E},i,k} - t_{{\rm E},k})^2/2\sigma^2_{t_{{\rm E},k}}]}{\sqrt{2\pi}\sigma_{t_{{\rm E},k}}},
\end{equation}
\begin{equation}
    \mathcal{L}_{i,k}(\thetae) = \frac{{\rm exp}[-(\theta_{{\rm E},i,k} - \theta_{{\rm E},k})^2/2\sigma^2_{\theta_{{\rm E},k}}]}{\sqrt{2\pi}\sigma_{\theta_{{\rm E},k}}},
\end{equation}
\begin{equation}
    \mathcal{L}_{i,k}(\bm{\pi}_{\rm E}) = \frac{{\rm exp}[-\sum_{m,n=1}^2b_{m,n}^k(\pi_{{\rm E},m,i}-\pi_{{\rm E},m,k})(\pi_{{\rm E},n,i}-\pi_{{\rm E},n,k})/2]}{2\pi/\sqrt{{\rm det}~b^k}},
\end{equation}
$b_{m,n}^k$ is the inverse covariance matrix of $\bm{\pi}_{{\rm E},k}$, and $(m,n)$ are dummy variables ranging over ($N, E$). Finally, we combine the Bayesian result of the $u_0 > 0$ and $u_0 < 0$ solutions by their Galactic-model likelihood and ${\rm exp}(-\Delta\chi^2_k/2)$, where $\Delta\chi^2_k$ is the $\chi^2$ difference between the $k$th solution and the best-fit solution.

The resulting posterior distributions of the host mass $M_{\rm host}$, the planet mass $M_{\rm planet}$, the lens distance $D_{\rm L}$ and the projected planet-host separation $a_\perp$ are listed in Table \ref{phy1} and shown in Figure \ref{baye}. The presented parameters are the median values of the Bayesian distributions, and the upper and lower limits correspond to the $15.9\%$ and $84.1\%$ percentages of their distributions, respectively. The Bayesian analysis yields a host mass of $M_{\rm host} = 0.61_{-0.24}^{+0.29}~M_{\odot}$, a planet mass of $M_{\rm planet} = 2.48_{-0.98}^{+1.19}~M_{\earth}$, a host-planet projected separation $a_\perp = 3.4_{-0.5}^{+0.5}$~au and a lens distance of $D_{\rm L} = 6.8_{-0.9}^{+0.6}$~kpc. The estimated physical parameters indicate that lens companion is a terrestrial planet located well beyond the snow line of the host star (assuming a snow line radius $a_{\rm SL} = 2.7(M/M_{\odot})$~{\rm au}, \citealt{snowline}). In addition, for an $M \sim 0.6~M_{\odot}$ star at a distance of $\sim 6.8$~kpc, it should be behind most of the dust extinction and its apparent magnitude should be $I_{\rm L} \sim 23$. Hence, it is estimated that the lens flux only contributes a very small fraction of the $I_{\rm B} \sim 19.4$ blended light.

We note that although the introduction of $\bm{\pi}_{\rm E}$ does not significantly improve the fit, it does constrain the amplitude of $\bm{\pi}_{\rm E, E}$ to be small, and thereby influences the mass estimate via Equation (\ref{eq:mass}). In particular, if we remove the $\bm{\pi}_{\rm E}$ term from Equation (\ref{eq:weight}), then the Bayesian host mass estimate is shifted lower to $M_{\rm host} = 0.52_{-0.30}^{+0.32}~M_{\odot}$. We also note that this is in good
agreement with the general prediction of \cite{KB190371}, for the case of $\thetae = 0.37$ mas and $\mu_{\rm rel} < 10~{\rm mas\,yr^{-1}}$ (and no other information), i.e., $M_{\rm host} = 0.45_{-0.23}^{+0.30}~M_{\odot}$. See their Figures 6 and 7.

%% file: dis.tex
\subsection{A New Path for the Mass-ratio Function}

For most microlensing planetary events, light-curve analyses do not provide the masses of the host and the planet, but the planet-host mass ratio, $q$, is well determined. There have been three studies about the microlensing planet-host mass-ratio function from homogeneous samples. \cite{mufun} adopted the 13 high-magnification events intensively observed by the Microlensing Follow Up Network ($\mu$FUN), which included six planets. \cite{Wise} used the 224 events observed by OGLE + MOA + Wise Observatory, including seven $q < 0.01$ planets. It confirmed the result of \cite{OB07368} that the planet occurrence rate increases while q decreases for $-4.5 < \log q < -2.0$. \cite{Suzuki2016} built a substantially larger sample that consisted of 1474 events discovered by the MOA-II microlensing survey alert system, the \cite{mufun} sample and 196 events from the PLANET follow-up network \citep{Cassan2012}, with 30 planets in total. This larger sample revealed a break in the mass-ratio function at about $q_{\rm break} = 17 \times 10^{-5}$, below which the planet occurrence rate decreases as $q$ decreases. 

KMT opens a window for the mass-ratio function down to $q \sim 10^{-5}$ and thus can test the break reported by \cite{Suzuki2016}. Including \eventLb, KMT has detected five very low mass-ratio planets whose mass ratios lie below the lowest mass ratio, $q = (4.43 \pm 0.029) \times 10^{-5}$ \citep{OB130341}, in the three samples mentioned above. The four other planets are KMT-2018-BLG-0029Lb with $q \sim 1.8 \times 10^{-5}$ \citep{KB180029}, KMT-2019-BLG-0842Lb with $q \sim 4.1 \times 10^{-5}$ \citep{KB190842}, OGLE-2019-BLG-0960Lb with $q \sim 1.4 \times 10^{-5}$ \citep{OB190960} and KMT-2020-BLG-0414Lb with $q \sim 1.1 \times 10^{-5}$ \citep{KB200414}. KMT data played a major or decisive role in all the five discoveries. However, it is challenging to build a homogeneous KMT sample, considering that there are $\sim 3000$ KMT events per year and the imperfect end-of-year-pipeline light curves. \cite{OB190960} proposed to construct a KMT high-magnification sample by placing a magnification threshold (e.g., $A_{\rm max} > 20$), but this approach would require intensive efforts on KMT TLC re-reductions. A second approach, proposed by \cite{KB200414}, is to systematically follow up high-magnification events in the KMT low-cadence ($\Gamma \lesssim 1~{\rm hr}^{-1}$) fields using Las Cumbres Observatory (LCO) global network and $\mu$FUN. Because the follow-up data would play a major role in the detections of planetary signals, this approach would require many fewer KMT TLC re-reductions (and so, much less effort) than the \cite{OB190960} approach, but it would require intensive effort to carry out the real-time monitoring and obtain follow-up observations.

The anomaly search to the KMT end-of-year-pipeline light curves provides a new path for the mass-ratio function with a large and homogeneous sample. This approach would only require KMT TLC re-reductions on candidate planetary events, and most of the KMT events can be included in the sample except a small fraction of events, e.g., events with a variable source. We applied the anomaly search to the known 2019 KMT planets, and all of them were identified as a candidate signal with the current search thresholds, including the two very low mass-ratio planets, KMT-2019-BLG-0842Lb with $\Delta\chi^2_{\rm zero} = 519$ and OGLE-2019-BLG-0960Lb with $\Delta\chi^2_{\rm zero} = 2623$. This should hold for almost all the 2016--2019 KMT planets\footnote{The 2020 season would not be considered due to Covid-19, for which two of KMT’s three observatories were shut down during most of the 2020 season.}, and the final planet sample from the 2016--2019 data should be at least two times larger than the \cite{Suzuki2016} sample. 

\subsection{Future Improvements of Anomaly Search}\label{future}

The main purpose of the current search is to develop and test the method and programming. The detection of the lowest mass-ratio planetary caustic to date illustrates the utility of this search. The ultimate goal of our search is to form a large and homogeneous sample to study the microlensing planet-host mass-ratio function down to $q \sim 10^{-5}$. To achieve it, the current search can be improved in several respects. 

First, the search could be extended to all of the 2016--2019 events without the current catalog-star brightness limit $\Icat < 19$. At present, only the 2019 data can be used, because the 2016--2018 data lack seeing and background information and the 2016--2017 end-of-year-pipeline light curves are not of sufficiently high quality. 

Second, the search could adopt shorter $\teff$ and lower $\chi^2_{\rm zero}$ thresholds. The lower limit of $\teff$ should be reduced to $\sim 0.05$ days, in order to find the shortest signals, at least in the $\Gamma \geq 4~{\rm hr}^{-1}$ fields, which cover $\sim 12~{\rm deg}^2$. Estimating that 10 points are required to characterize a short anomaly, the detection threshold for these high cadence fields is  $t_{\rm eff,limit} \sim (10/\Gamma)/2 = 0.05$ days. For the planetary signal of \event, its best-fit has $\teff \sim 0.1$ days, with $\Delta\chi^2 = 49$ better than the model with $\teff = 0.3$ days. The disadvantage is that decreasing the lower limit of $\teff$ leads to many more anomaly candidate signals that must be reviewed by the operator. Using $t_{{\rm eff},1} = 0.05$ days, and $\Delta\chi^2_{\rm zero} = 50$ and $\Delta\chi^2_{\rm flat} = 20$ as the thresholds, the anomaly search to the current 997-event sample yields 15486 candidate signals from 511 events. Thus, it should have about 40000 signals for one season of events and take the operator about 50 hours to review them, which is acceptable.  


Third, it is important to form a review and modeling group. The group would significantly reduce the bias of one operator and avoid missing signals. In addition, there would be about 200 anomalous events per year. Although most of these events are not planetary events, considerable modeling would be required to identify all of the planets.

\acknowledgments
W.Z., S.M. and X.Z. acknowledge support by the National Science Foundation of China (Grant No. 11821303 and 11761131004). This research has made use of the KMTNet system operated by the Korea Astronomy and Space Science Institute (KASI) and the data were obtained at three host sites of CTIO in Chile, SAAO in South Africa, and SSO in Australia. The OGLE has received funding from the National Science Centre, Poland, grant MAESTRO 2014/14/A/ST9/00121 to AU. The MOA project is supported by JSPS KAK-ENHI Grant Number JSPS24253004, JSPS26247023, JSPS23340064, JSPS15H00781, JP16H06287, JP17H02871 and JP19KK0082. Work by JCY was supported by JPL grant 1571564. Work by C.H. was supported by the grants of National Research Foundation of Korea (2019R1A2C2085965 and 2020R1A4A2002885). This research has made use of the NASA Exoplanet Archive, which is operated by the California Institute of Technology, under contract with the National Aeronautics and Space Administration under the Exoplanet Exploration Program. 

\software{pySIS \citep{pysis}, pyDIA (Zenodo, \url{doi:10.5281/zenodo.268049}, as developed on GitHub), OGLE DIA pipeline \citep{Wozniak2000}, \Sp\ photometry software \citep{Spitzerdata}}

\appendix

\section{Analysis Including \Sp\ Data}
\label{sec:spitzer}


Simultaneously observing the same microlensing event from Earth and one well-separated satellite \citep{1966MNRAS.134..315R} can yield the measurements of satellite microlens parallax (see Figure 1 of \citealt{1994ApJ...421L..75G}),
\begin{equation}\label{equ:para}
\vec{\pi}_{\rm E} = \frac{\rm au}{D_{\perp}}\left(\Delta\tau,\Delta\beta\right),
\end{equation}
with, e.g., 
\begin{equation}\label{equ:para2}
\Delta\tau\equiv\frac{t_{0,Spitzer}-t_{0,\rm\oplus}}{t_{\rm E}}; \qquad \Delta\beta\equiv\pm u_{0,Spitzer}-\pm u_{0,\oplus},
\end{equation}
where $D_\perp$ is the projected separation between the \Sp\ satellite and Earth at the time of the event. 

\event\ was selected as a ``secret'' target for \Sp\ observations on 14 July 2019 and was formally announced as a ``Subjective, immediate'' (SI) \Sp\ target on 18 July 2019. The goal of the \Sp\ microlensing program is to create an unbiased sample of microlensing events with well-measured parallax for measuring the Galactic distribution of planets in different stellar environments \citep{Novati2015,Zhu2017spitzer}. See \citet{YeeSpitzer} for the detailed protocols for the selection and observational cadence of \Sp\ targets. The \Sp\ observations began on 20 July 2019 (${\rm HJD}^{\prime} = 8685.1$) and ended on 16 August 2019 (${\rm HJD}^{\prime} = 8712.0$), with 22 data points in total. Each \Sp\ observation was composed of six dithered 30s exposures using the 3.6 $\mu$m channel ($L-$band) of the IRAC camera. The \Sp\ data were reduced by the method presented by \cite{Spitzerdata}.

The \Sp\ light curve shown in Figure \ref{lc2} exhibits a steady decline during the \Sp\ observing window. The first \Sp\ observation is at 8685.1, whereas the peak of the light curve as seen from the ground is at 8690.6. This implies $\pi_{\rm E, E} \gtrsim 0$. In addition,  we include a $VIL$ color-color constraint on the \Sp\ source flux by matching the OGLE-III and \Sp\ photometry for red-giant stars within $1'$ and find
\begin{equation}\label{VIL}
   (I_{\rm OGLE} - L_{Spitzer}) = 1.675 \pm 0.042.
\end{equation}

To compare the satellite microlens parallax with the ground-based parallax, we first fit for the {\it Spitzer}-``ONLY'' parallax \citep{OB180596} using the method of \cite{KB180029}. We fix ($t_0, u_0, \te, s, q, \alpha, \rho$) along with the $I$-band source flux as the best-fit parameters for the 2L1S ground-based parallax models and then derive a grid of ($\pi_{\rm E,N}, \pi_{\rm E,E}$) with a spacing of 0.005. We repeat the analysis for both the $u_0 > 0$ and $u_0 < 0$ solutions. The resulting parallax contours are shown in the upper panels of Figure \ref{pie}. The form of the {\it Spitzer}-``ONLY'' contours is intermediate between the four-fold degeneracy predicted by \cite{1966MNRAS.134..315R} (and illustrated in Figure 1 of \citealt{1994ApJ...421L..75G}) and the arc-like contours analyzed by \cite{Gould2019} for the case of late-time, monotonically declining, observations. That is, for each case ($u_{0,\oplus} > 0$ and $u_{0,\oplus} < 0$), there are two distinct solutions at the $1\sigma$ level, but these are connected by arcs at the $2\sigma$ level. See \cite{Gould2019} for a discussion of these transition-contour morphologies. Figure \ref{pie} shows that the $2\sigma$ contours of the {\it Spitzer}-``ONLY'' parallax overlap the $1\sigma$ contour of the ground-based parallax (and vice versa), so there is no tension between the two parallax constraints. We then fit the full-parallax models by combining the ground-based and \Sp\ data. The resulting parallax contours are shown in the lower panels of Figure \ref{pie}. For both $u_0 > 0 $ and $u_0 < 0$, the arc-like {\it Spitzer}-``ONLY'' parallax is broken into two discrete minima due to the ``1-D'' constraint of ground-based parallax. We label the four discrete minima in total by (``$u_0 > 0$ \& small $\pie$'', ``$u_0 > 0$ \& large $\pie$'', ``$u_0 < 0$ \& small $\pie$'', ``$u_0 < 0$ \& large $\pie$'') and present their lensing parameters in Table \ref{parm2}. We find that the non-parallax parameters of the full-parallax models are consistent with the parameters of the static model at $1\sigma$. 

Before making a detailed Bayesian analysis, we can roughly estimate the physical parameters and compare to the Bayesian results from the ground-based data. First, using the results of \cite{Gould2020}, we can see that the smaller-parallax local minima are strongly favored. His Equation (15) states that the relative probability of two isolated minima with equal $\chi^2$ from the light curve is given by
\begin{equation}\label{eqn:gould20}
P \propto \rho(D_{\rm L}) D_{\rm L}^4 \pi_{\rm E}^{-1} \Phi(M) f(\bm{\mu}),
\end{equation}
where $\rho(D_L)$ is the local density, $\Phi(M)$ is the mass function and $f(\bm{\mu})$ is the 2-D relative proper-motion distribution at $D_{\rm L}$. For the two solutions in each of the two panels of Figure \ref{pie}, the parallaxes are $\pie \sim 0.1$ and 0.4, the distances are $D_{\rm L} \sim 6.5$ and 4.0 kpc, and the densities are in a ratio of about 10 : 1. Hence, the combined ratios of the first three terms of Equation (\ref{eqn:gould20}) are $10 \times (6.5/4)^4 \times 4 \sim 260$. This factor overwhelms the $\chi^2 \sim 2$ advantage of the large-parallax solution as well as the slight differences in the last two factors. Second, combining $\pie \sim 0.1$ and $\thetae \sim 0.37$, the lens system should have $M_{\rm L} \sim 0.45$ and $D_{\rm L} \sim 6.5$ kpc.  

Finally, we repeat the Bayesian analysis for the full-parallax models and show the resulting physical parameters in Table \ref{phy2}. The results are quite consistent with the estimates above. In addition, \citet{Zhu2017spitzer} proposed that events should have
\begin{equation}
    \sigma(D_{8.3}) < 1.4~{\rm kpc}; \qquad D_{8.3} \equiv \frac{\rm kpc}{1/8.3 + \pi_{\rm rel}/{\rm mas}}
\end{equation}
to be included in the \Sp\ statistical sample. We follow the methods of \cite{OB161190} to fit with a PSPL model using the analogous data and conduct a Bayesian analysis without the $\thetae$ weight. We find $\sigma(D_{8.3}) = 0.7$ kpc, and thus OGLE-2019-BLG-1053Lb can be included in the statistical sample of \Sp\ events if the systematics of \Sp\ data does not affect the parallax measurements. However, the total flux change of the \Sp\ light curve is $\sim 2.0$ instrumental flux unit, which is only a few times the level of systematics seen in other \Sp\ events with observations in the baseline \citep{KB180029,OB170406,OB180799}. We therefore leave the question of whether OGLE-2019-BLG-1053Lb can be included in the final \Sp\ statistical sample to a future comprehensive analysis of \Sp\ planets. Here we simply note that, because the ``SI'' observing decision was made a week before the planetary anomaly, it was not influenced in any way by the presence of a planet, thereby satisfying a key criterion of \cite{YeeSpitzer}. Indeed, \event\ is the second example (after KMT-2018-BLG-0029, \citealt{KB180029}) of a very low-$q$ planetary event observed by \Sp\ for which the planet remained unnoticed until well after the end of the season. This fact makes clear the need for an intensive review of all \Sp\ events.

%% file: table.tex
\begin{table}[ht]
    \renewcommand\arraystretch{1.2}
    \centering
    \caption{Data used in the analysis with corresponding data reduction method and rescaling factors}
    \begin{tabular}{c c c c c c c c c}
    \hline
    \hline
    Collaboration & Site & Filter & Coverage (${\rm HJD}^{\prime}$) & $N_{\rm data}$ & Reduction Method & $k$ & $e_{\rm min}$ \\
    \hline
    OGLE &  & $I$ &  8521.9 -- 8787.5 & 811 & \cite{Wozniak2000} & 1.400 & 0.011 \\
    KMTNet & SSO (03) & $I$ & 8534.3 -- 8777.9 & 1635 & pySIS$^1$ & 1.506 & 0.000 \\
    KMTNet & SSO (43) & $I$ & 8534.3 -- 8777.9 & 1627 & pySIS & 1.375 & 0.000 \\
    KMTNet & CTIO (03) & $I$ & 8533.8 -- 8777.5 & 2050 & pySIS & 1.207 & 0.000 \\
    KMTNet & CTIO (43) & $I$ & 8533.9 -- 8775.5 & 2011 & pySIS & 1.136 & 0.000 \\
    KMTNet & SAAO (03) & $I$ & 8536.6 -- 8777.3 & 1783 & pySIS & 1.335 & 0.000 \\
    KMTNet & SAAO (43) & $I$ & 8537.6 -- 8777.3 & 1782 & pySIS & 1.499 & 0.000 \\
    \Sp\   &           & $L$ & 8685.1 -- 8712.0 & 22 & \cite{Spitzerdata} & 2.64 & 0.000 \\
    \hline
    KMTNet & CTIO (03) & $I$ & 8533.8 -- 8777.5 & 2050 & pyDIA$^2$ &  &  \\
    KMTNet & CTIO (03) & $V$ & 8533.9 -- 8768.5 & 200  & pyDIA &  &  \\    
    \hline
    \hline
    \end{tabular}
    \tablecomments{${\rm HJD}^{\prime} = {\rm HJD} - 2450000$.\\
    $^1$ \cite{pysis} \\
    $^2$ MichaelDAlbrow/pyDIA: Initial Release on Github, doi:10.5281/zenodo.268049\\
    }
    \label{data}
\end{table}

\begin{table*}[htb]
    \renewcommand\arraystretch{1.25}
    \centering
    \caption{Parameters for PSPL and 2L1S models using Ground-based data}
    \begin{tabular}{c c c c c c}
     \hline
    Parameter & PSPL & \multicolumn{2}{c}{2L1S Static} & \multicolumn{2}{c}{2L1S Parallax} \\
    \hline
      &  & Close & Wide & Wide $u_0 > 0$ & Wide $u_0 < 0$ \\
    $\chi^2/dof$ & 12130.6/11682 & 11718.6/11678 & 11677.0/11678 & 11676.1/11676 & 11675.8/11676 \\
    \hline
    $t_{0}$ (${\rm HJD}^{\prime}$) & $8690.462 \pm 0.040$  & $8690.538 \pm 0.042$ & $8690.555 \pm 0.044$ & $8690.572 \pm 0.049$ & $8690.566 \pm 0.055$  \\
    $u_{0}$  & $0.373 \pm 0.016$ & $0.355 \pm 0.011$  & $0.350 \pm 0.010$ & $0.352 \pm 0.013$ & $-0.350 \pm 0.011$ \\
    $\te$  & $32.8 \pm 1.0$ & $33.7 \pm 0.7$ & $34.1 \pm 0.7$ & $34.3 \pm 1.0$ & $34.4 \pm 1.0$ \\
    $s$ & ... & $0.707 \pm 0.006$ & $1.406 \pm 0.011$ & $1.407 \pm 0.013$ & $1.406 \pm 0.011$ \\
    $q (10^{-5})$ & ... & $3.14 \pm 0.30$ & $1.29 \pm 0.10$ & $1.25 \pm 0.12$ & $1.24 \pm 0.13$ \\
    $\alpha$ (rad) & ... & $0.507 \pm 0.005$ & $3.664 \pm 0.004$ & $3.683 \pm 0.026$ & $-3.681 \pm 0.026$ \\
    $\rho (10^{-3})$ & ... & $2.54 \pm 0.58$ & $2.19 \pm 0.16$ & $2.08 \pm 0.18$ & $2.07 \pm 0.22$ \\
    $\pi_{\rm E, N}$ & ... & ... & ... & $0.338 \pm 0.475$ & $-0.327 \pm 0.515$ \\
    $\pi_{\rm E, E}$ & ... & ... & ... & $-0.012 \pm 0.089$ & $0.027 \pm 0.053$ \\
    $I_{\rm S}$ & $19.797 \pm 0.064$ & $19.865 \pm 0.045$ & $19.886 \pm 0.043$ & $19.877 \pm 0.051$ & $19.888 \pm 0.044$ \\
    $I_{\rm B}$ & $19.429 \pm 0.044$ & $19.383 \pm 0.028$ & $19.370 \pm 0.026$ & $19.377 \pm 0.029$ & $19.368 \pm 0.026$ \\
    \hline
    \end{tabular}
    \tablecomments{${\rm HJD}^{\prime} = {\rm HJD} - 2450000$}
    \label{parm1}
\end{table*}   

\begin{table}[htb]
    \renewcommand\arraystretch{1.5}
    \centering
    \caption{Physical parameters for \event\ using Ground-based data}
    \begin{tabular}{c|c c c c c| c c}
    \hline
    \hline
     Solutions & \multicolumn{5}{c|}{Physical Properties} & \multicolumn{2}{c}{Relative Weights} \\
    \hline
     & $M_{\rm host}[M_{\odot}]$ & $M_{\rm planet}[M_{\oplus}]$ & $D_{\rm L}$[kpc] & $a_{\perp}$[au] & $P_{\rm bulge}$ & Gal.Mod. & $\chi^2$ \\
    \hline
    $u_0 > 0$ & $0.60_{-0.24}^{+0.29}$ &  $2.46_{-1.00}^{+1.21}$ & $6.7_{-1.0}^{+0.6}$ &  $3.4_{-0.5}^{+0.5}$ & 0.672 & 1.000 & 0.861 \\

    $u_0 < 0$ & $0.62_{-0.23}^{+0.28}$ &  $2.50_{-0.96}^{+1.18}$ & $6.8_{-0.8}^{+0.6}$ & $3.4_{-0.5}^{+0.4}$ & 0.718 & 0.894 & 1.000 \\
    
    Total & $0.61_{-0.24}^{+0.29}$ &    $2.48_{-0.98}^{+1.19}$ & $6.8_{-0.9}^{+0.6}$ & $3.4_{-0.5}^{+0.5}$ & 0.695 &  & \\
    \hline
    \hline
    \end{tabular}
    \tablecomments{$P_{\rm bulge}$ is the probability of a lens system in the Galactic bulge. The combined result is obtained by a combination of $u_0 > 0$ and $u_0 < 0$ solutions weighted by the probability for the Galactic model (Gal.Mod.) and ${\rm exp}(-\Delta\chi^2/2)$.}
    \label{phy1}
\end{table}

\begin{table*}[htb]
    \renewcommand\arraystretch{1.25}
    \centering
    \caption{Parameters for 2L1S models using Ground-based + \Sp\ data}
    \begin{tabular}{c c c c c}
     \hline
    Parameter & $u_0 > 0$ \& small $\pie$ & $u_0 > 0$ \& large $\pie$ & $u_0 < 0$ \& small $\pie$  & $u_0 < 0$ \& large $\pie$ \\
    \hline
    $\chi^2/dof$  & 11698.5/11696 & 11696.3/11696 & 11700.0/11696 & 11698.3/11696 \\
    \hline
    $t_{0}$ (${\rm HJD}^{\prime}$)  & $8690.615 \pm 0.044$ & $8690.586 \pm 0.044$ & $8690.610 \pm 0.045$ & $8690.603 \pm 0.045$  \\
    $u_{0}$  & $0.351 \pm 0.009$ & $0.351 \pm 0.008$ & $-0.350 \pm 0.006$ & $-0.346 \pm 0.007$ \\
    $\te$  & $34.0 \pm 0.6$ & $33.5 \pm 0.5$ & $34.1 \pm 0.4$ & $33.9 \pm 0.5$ \\
    $s$ & $1.406 \pm 0.009$ & $1.407 \pm 0.008$ & $1.404 \pm 0.006$ & $1.401 \pm 0.007$ \\
    $q (10^{-5})$ & $1.26 \pm 0.09$ & $1.36 \pm 0.09$ & $1.26 \pm 0.09$ & $1.31 \pm 0.19$ \\
    $\alpha$ (rad) & $3.664 \pm 0.004$ & $3.643 \pm 0.003$ & $-3.664 \pm 0.003$ & $-3.642 \pm 0.004$ \\
    $\rho (10^{-3})$ & $2.11 \pm 0.14$ & $2.24 \pm 0.16$ & $2.11 \pm 0.13$ & $2.18 \pm 0.14$ \\
    $\pi_{\rm E, N}$ & $-0.027 \pm 0.010$ & $-0.419 \pm 0.014$ & $0.027 \pm 0.010$ & $0.416 \pm 0.011$ \\
    $\pi_{\rm E, E}$ & $0.114 \pm 0.031$ & $0.124 \pm 0.031$ & $0.110 \pm 0.029$ & $0.096 \pm 0.032$ \\
    $I_{\rm S}$ & $19.882 \pm 0.036$ & $19.881 \pm 0.032$ & $19.889 \pm 0.023$ & $19.904 \pm 0.028$ \\
    $I_{\rm B}$ & $19.372 \pm 0.023$ & $19.373 \pm 0.020$ & $19.368 \pm 0.014$ & $19.359 \pm 0.017$ \\
    \hline
    \end{tabular}
    \tablecomments{${\rm HJD}^{\prime} = {\rm HJD} - 2450000$; $u_{0}$ is the impact parameter see from ground.}
    \label{parm2}
\end{table*}   

\begin{table}[htb]
    \renewcommand\arraystretch{1.5}
    \centering
    \caption{Physical parameters for \event\ using Ground-based + \Sp\ data}
    \begin{tabular}{c|c c c c c| c c}
    \hline
    \hline
     Solutions & \multicolumn{5}{c|}{Physical Properties} & \multicolumn{2}{c}{Relative Weights} \\
    \hline
     & $M_{\rm host}[M_{\odot}]$ & $M_{\rm planet}[M_{\oplus}]$ & $D_{\rm L}$[kpc] & $a_{\perp}$[au] & $P_{\rm bulge}$ & Gal.Mod. & $\chi^2$ \\
    \hline
     $u_0 > 0$ & $0.49_{-0.13}^{+0.18}$ &   $1.99_{-0.54}^{+0.79}$ & $6.6_{-0.7}^{+0.6}$ &  $3.3_{-0.4}^{+0.4}$ & 0.700 & 0.780 & 1.000 \\

     $u_0 < 0$ & $0.45_{-0.13}^{+0.18}$ &  $1.83_{-0.56}^{+0.78}$ & $6.4_{-0.8}^{+0.7}$ & $3.2_{-0.5}^{+0.4}$ & 0.543 & 1.000 & 0.369 \\
    
     Total & $0.48_{-0.13}^{+0.18}$ &   $1.93_{-0.56}^{+0.79}$ & $6.5_{-0.8}^{+0.6}$ & $3.3_{-0.4}^{+0.4}$ & 0.650 &  & \\    
    \hline
    \hline
    \end{tabular}
    \tablecomments{$P_{\rm bulge}$ is the probability of a lens system in the Galactic bulge. The combined result is obtained by a combination of $u_0 > 0$ and $u_0 < 0$ solutions weighted by the probability for the Galactic model (Gal.Mod.) and ${\rm exp}(-\Delta\chi^2/2)$.}
    \label{phy2}
\end{table}

%% file: figure.tex
\begin{figure}[htb] 
    \centering
    \includegraphics[width=0.70\columnwidth]{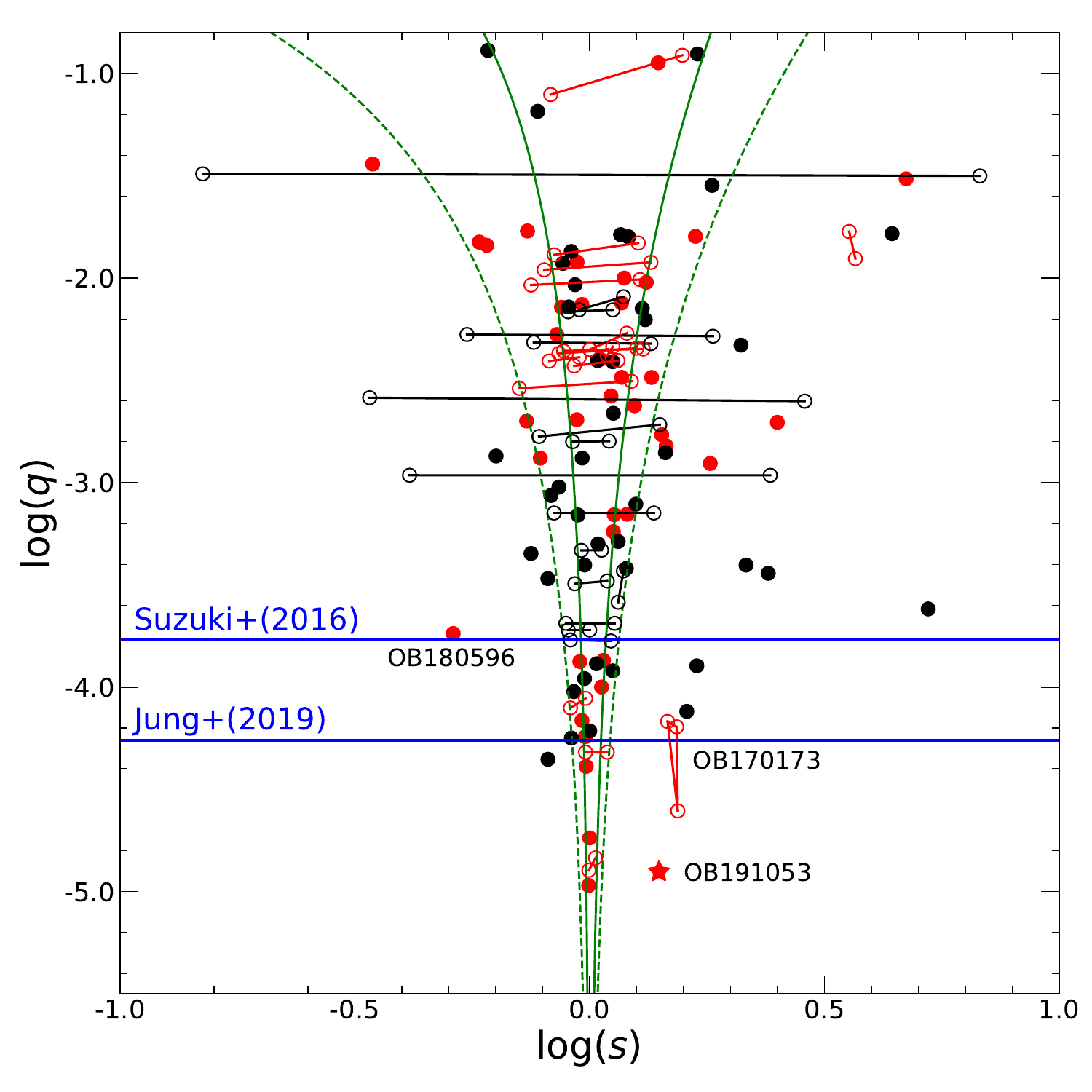}
    \caption{Microlensing parameters $(\log s,\log q)$ for planetary events, adapted from Figure 11 of \cite{OB190960}. The black and red points represent planets detected with and without KMTNet data, respectively. The red asterisk is the planet OGLE-2019-BLG-1053Lb found by the systematic search presented in this paper. Solutions are considered to be ``unique'' (filled points) if there are no competing solutions within $\Delta\chi^2 < 10$. Otherwise, they are shown by pairs of open circles linked by a line segment. There are eight such pairs for which $q$ differs by more than a factor of two. Seven of these are excluded on the grounds that $q$ is not accurately measured, but OGLE-2017-BLG-0173 \citep{OB170173} is preserved because it was detected by a channel of pure-planetary caustic and all of its degenerate solutions have $\log q < -4$. The three $\log q < -3$ planets detected with KMTNet data are marked with text. The power-law ``breaks'' proposed by \cite{Suzuki2016} and \citet{KB170165} are indicated with the blue lines. The two green solid lines represent the boundaries between resonant and non-resonant caustics using the Equation (59) of \cite{Dominik1999}, and the two green dashed lines show the boundaries for ``near-resonant'' caustics proposed by \cite{OB190960}. }
    \label{qs}
\end{figure}

\begin{figure}[htb] 
    \centering
    \includegraphics[width=0.63\columnwidth]{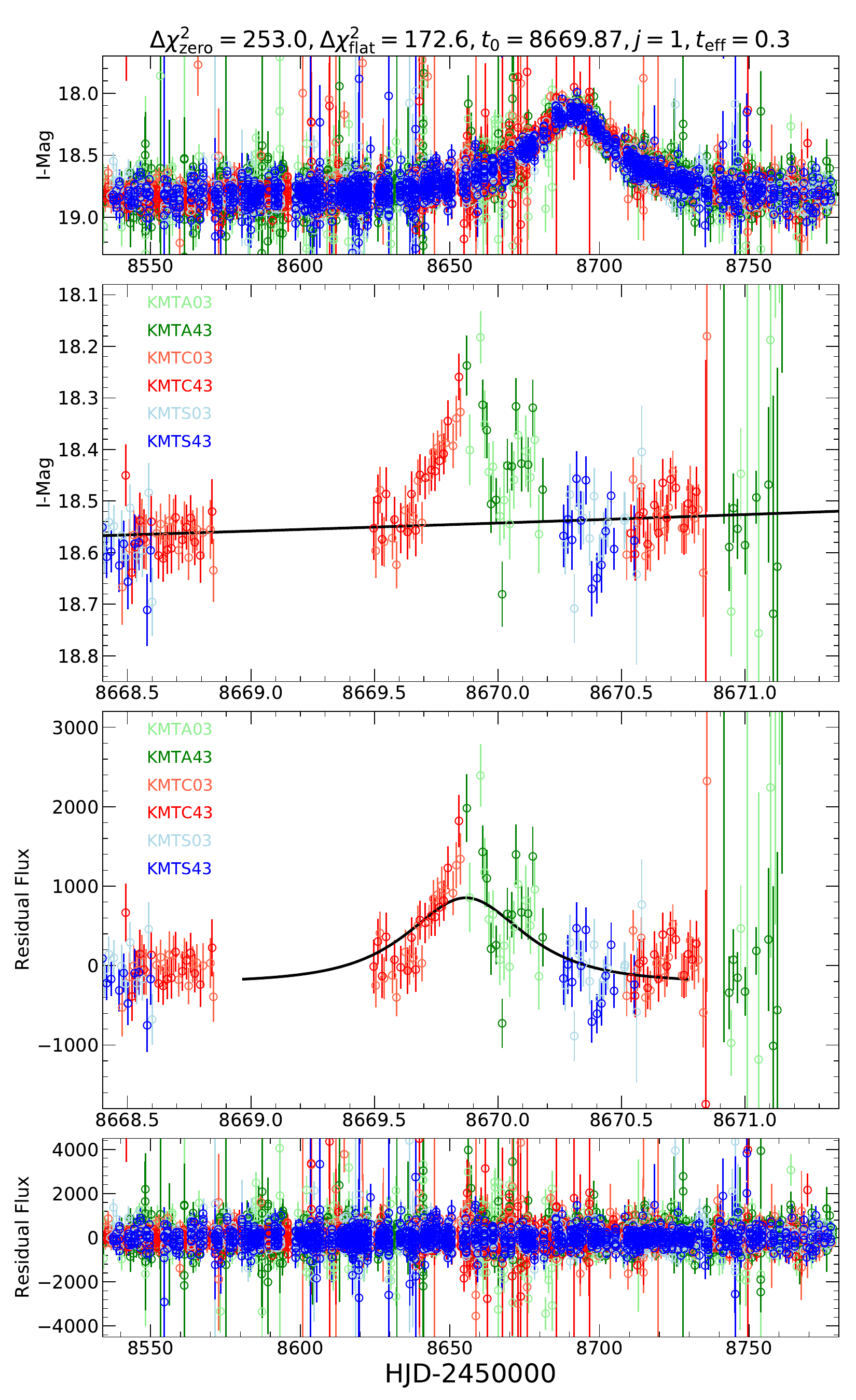}
    \caption{Example of the candidate signal of \event\ (ultimately judged to be real) as shown to the operator. The first and fourth panels show the whole season of data and their residuals to the PSPL model, respectively. The second and third panels show a zoom ($t_0 \pm 5~\teff$) of the candidate signal. The circles with different colors are observed data points for different data sets. The black line in the second panel represents the best-fit PSPL model, and the black line in the third panel represents the best-fit grid-search model for $t_0 \pm 3~\teff$. Five parameters are shown above the first panel: $\Delta\chi^2_{\rm zero}$, $\Delta\chi^2_{\rm flat}$, $t_0$, $j$ and $\teff$.}
    \label{review}
\end{figure}

\begin{figure}[htb] 
    \centering
    \includegraphics[width=0.85\columnwidth]{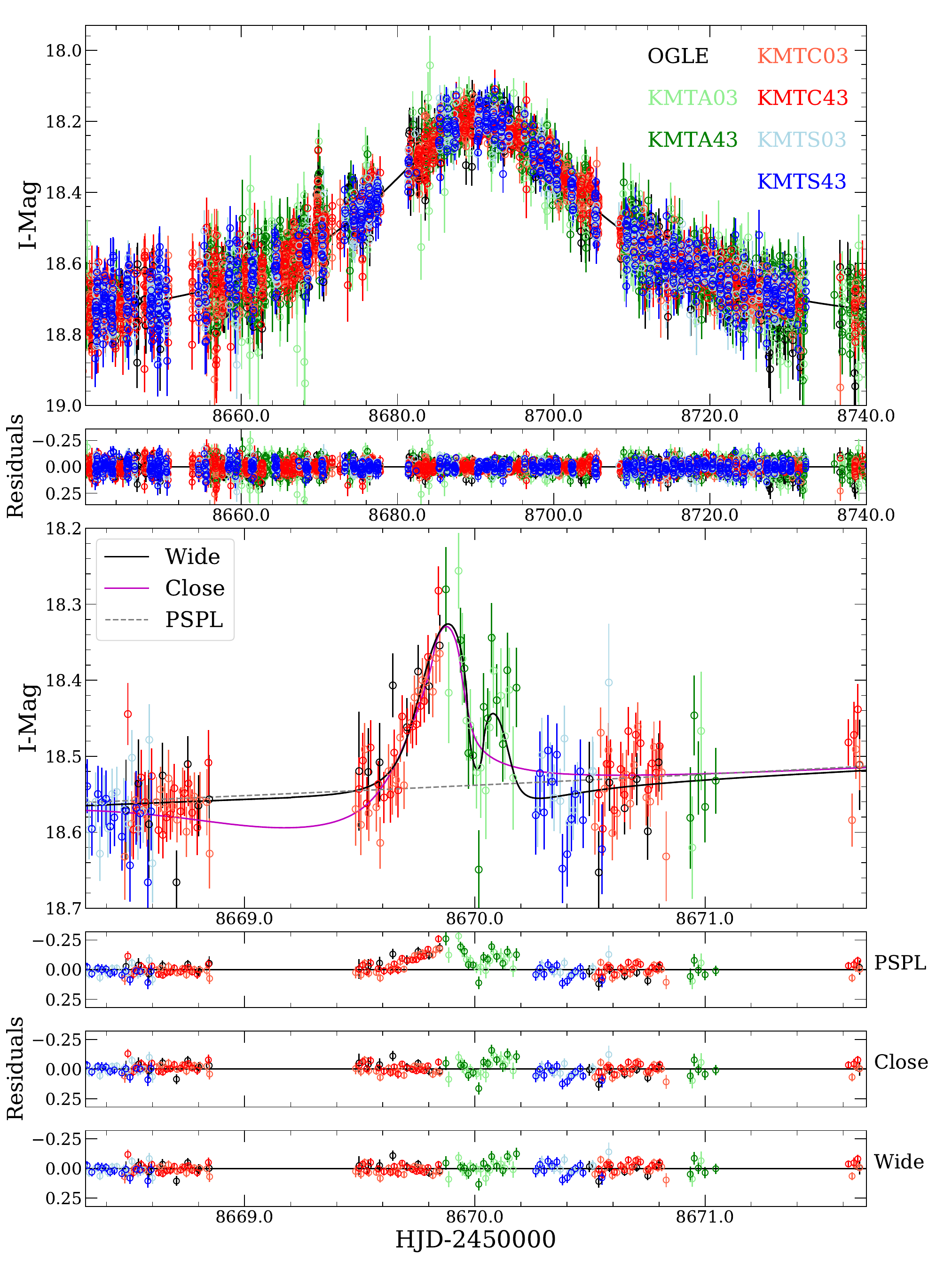}
    \caption{The observed data with the best-fit models. The open circles with different colors are observed ground-based data points for different data sets. The bottom four panels show a close-up of the planetary signal and the residuals to different models. The black and magenta solid lines represent the best-fit 2L1S Wide and Close models, respectively, and the black dashed line represents the best-fit PSPL model.}
    \label{lc}
\end{figure}

\begin{figure}[htb] 
    \centering
    \includegraphics[width=0.63\columnwidth]{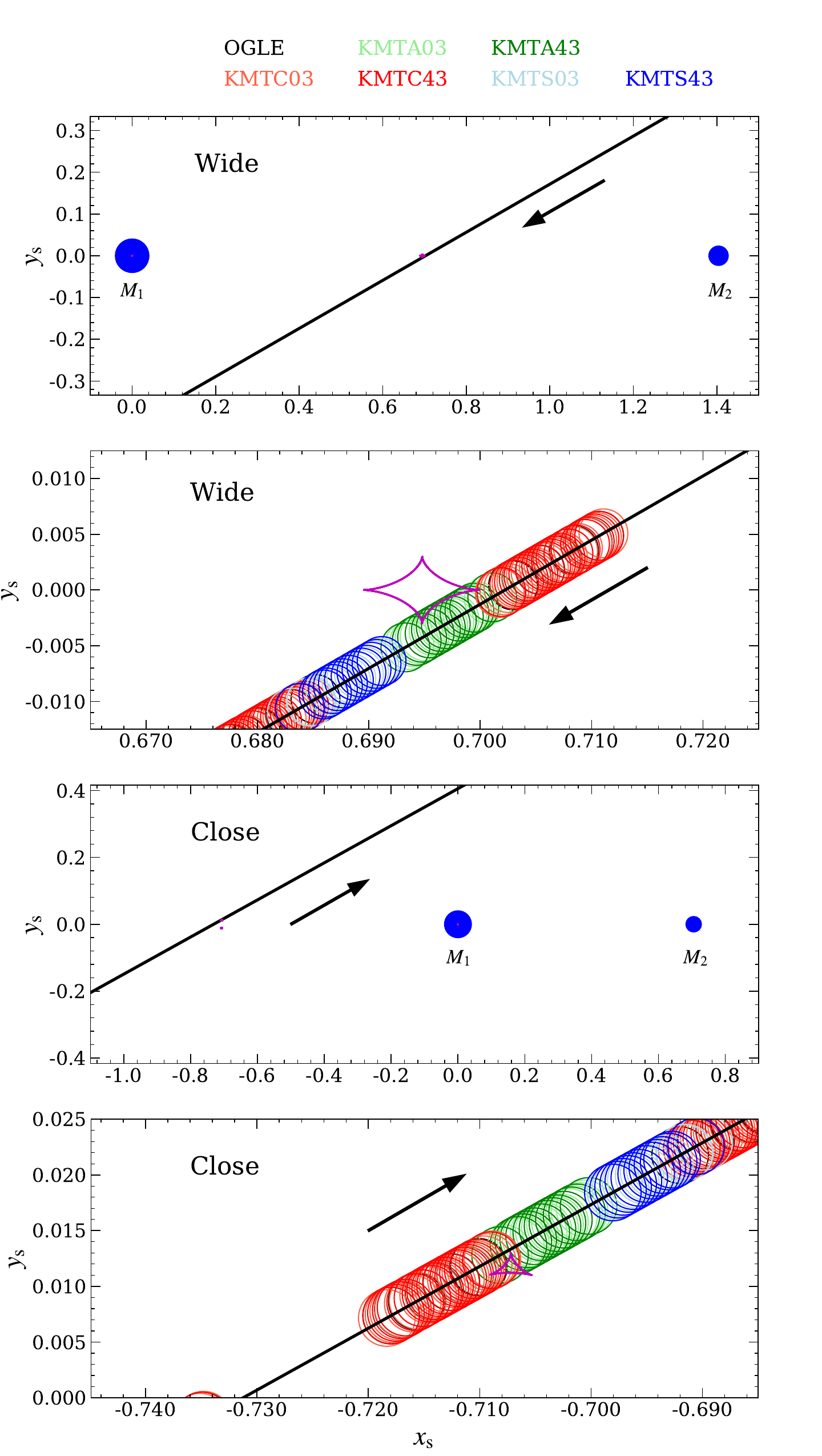}
    \caption{Geometries of the 2L1S Wide and Close models. In each panel, the magenta lines represent the caustic structure, the black solid line is the trajectory of the source, and the arrow indicates the direction of the source motion. The open circles with different colors represent the source location at the times of observation from different telescopes. The radii of the circles represent the best-fit source radius $\rho$. The blue dots, marked by $M_1$ (host) and $M_2$ (planet) are the positions of the two components of the lens.}
    \label{cau}
\end{figure}

\begin{figure}[htb] 
    \centering
    \includegraphics[width=0.70\columnwidth]{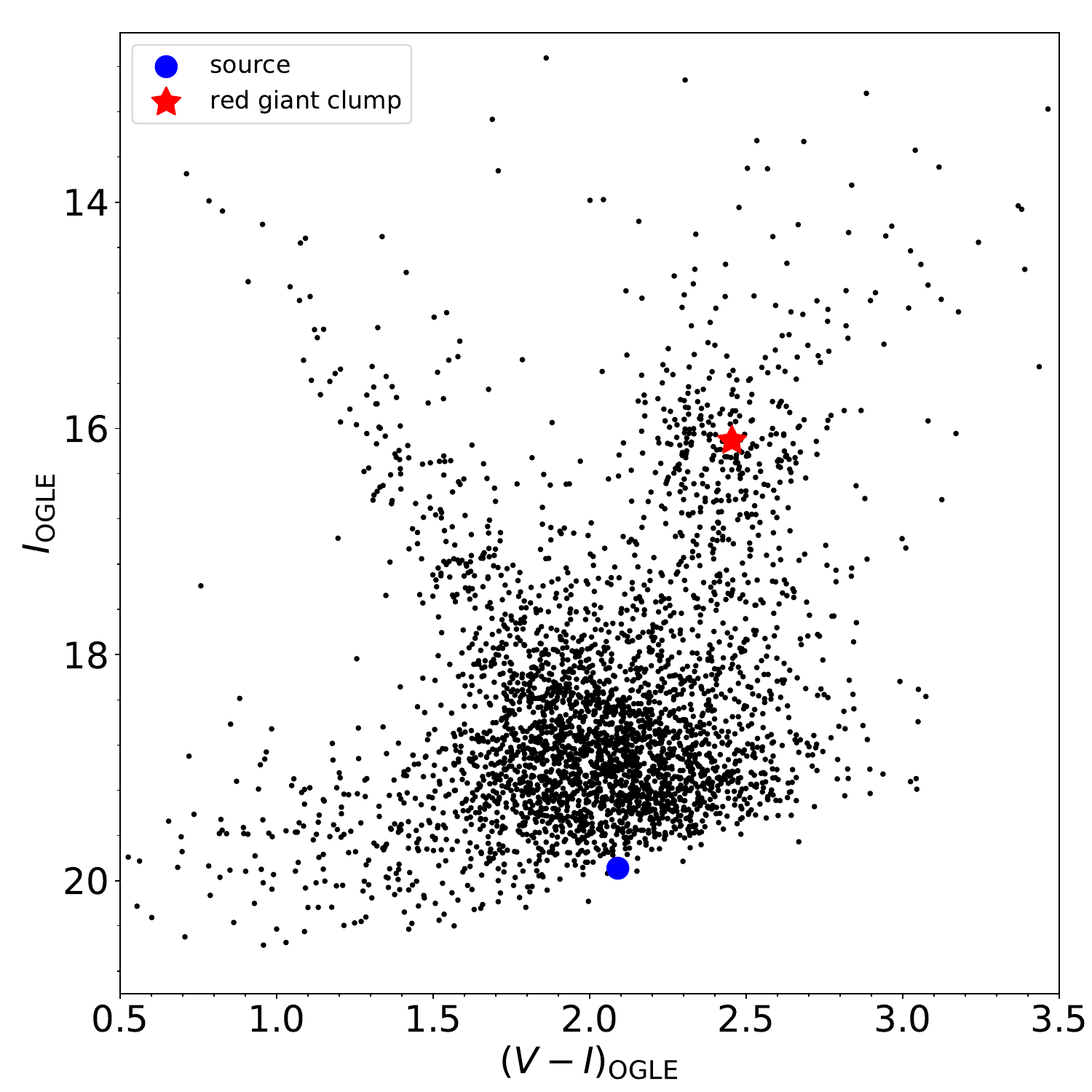}
    \caption{Color-magnitude diagram (CMD) for field stars within $80^{\prime\prime}$ centered on \event\ using the OGLE-III star catalog \citep{OGLEIII}. The red asterisk and blue dot represent the positions of the centroid of the red giant clump and the microlensing source star, respectively.}
    \label{cmd}
\end{figure}

\begin{figure}[htb] 
    \centering
    \includegraphics[width=0.70\columnwidth]{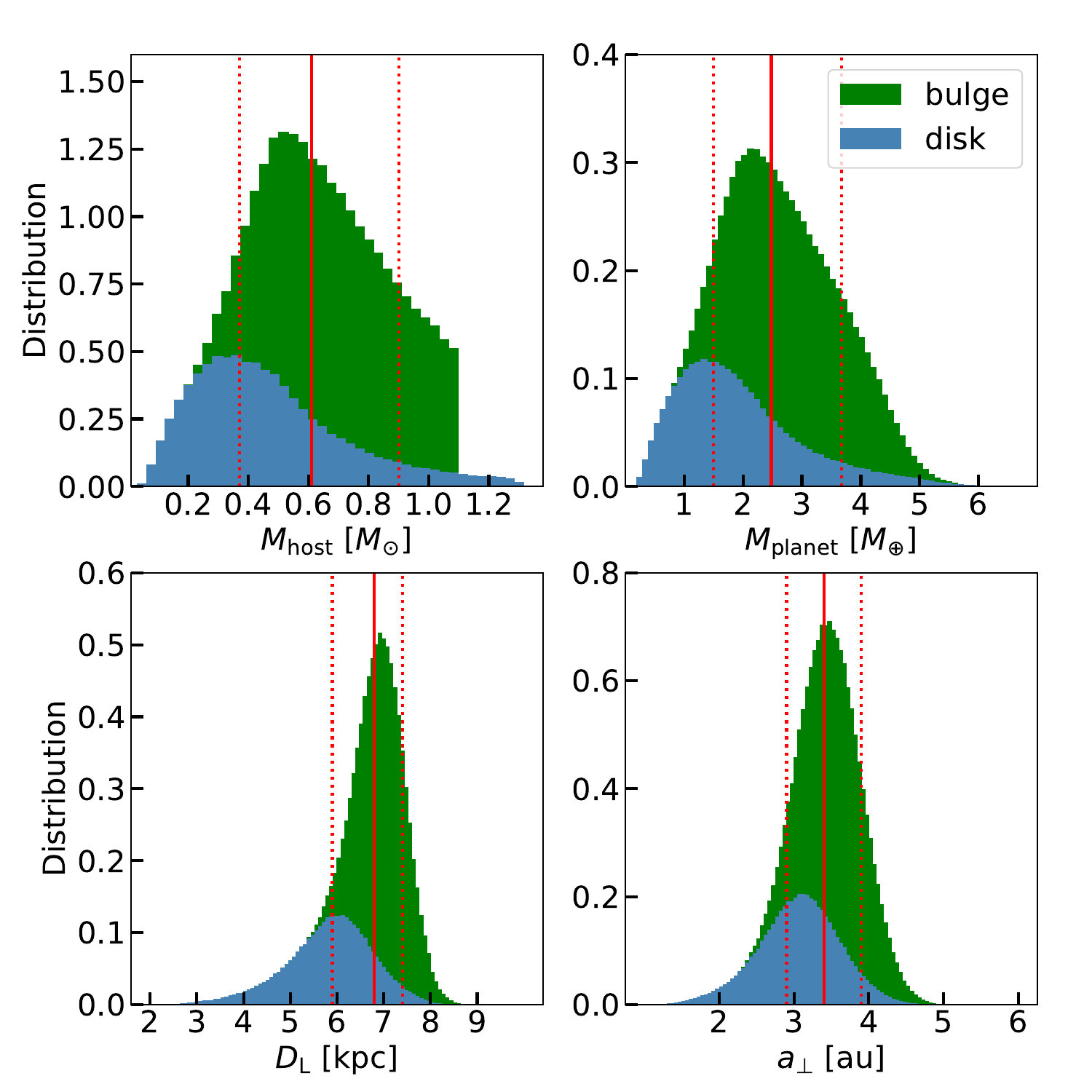}
    \caption{Bayesian posterior distributions of the lens mass $M_{\rm L}$, the planet mass $M_{\rm planet}$, the lens distance $D_{\rm L}$ and the projected planet-host separation $a_\perp$. The distributions are the combined results of the $u_0 > 0$ and $u_0 < 0$ solutions by their Galactic-model likelihood and ${\rm exp}(-\Delta\chi^2/2)$, where $\Delta\chi^2$ is the $\chi^2$ difference between the two solutions. In each panel, the red solid vertical line and the two red dashed lines represent the median value and the 15.9\% and 84.1\% percentages of the distribution. The total distribution is divided into bulge (green) and disk (blue) lenses. The upper limits of the host mass is $1.1~M_{\odot}$ for bulge lenses and $1.3~M_{\odot}$ for disk lenses. See Section \ref{Baye}.}
    \label{baye}
\end{figure}

\begin{figure}[htb] 
    \centering
    \includegraphics[width=0.85\columnwidth]{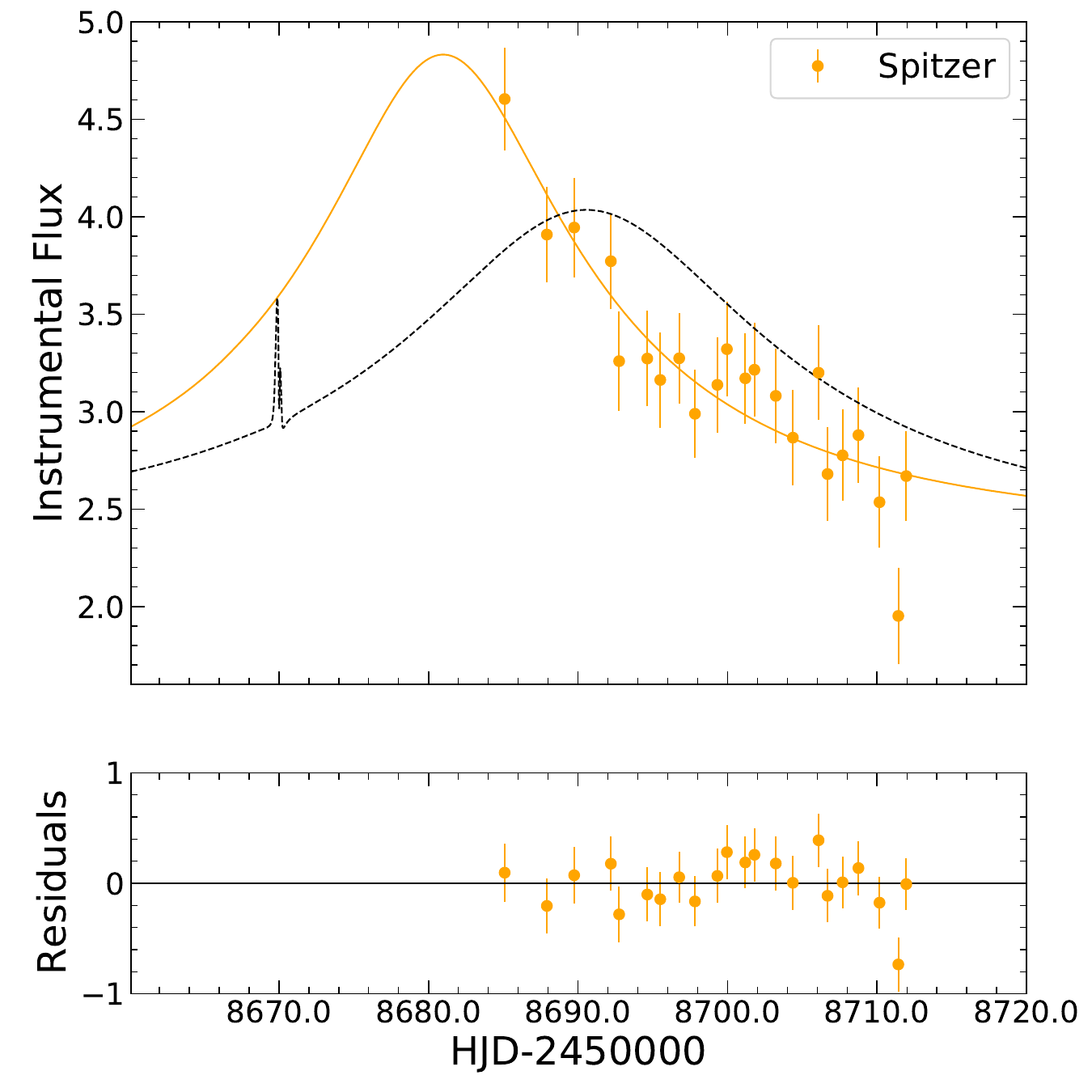}
    \caption{The observed \Sp\ data (orange points) in instrumental-flux units and the best-fit {\it Spitzer}-``ONLY'' model (the orange solid curve). The black dotted line represents the \Sp\ flux predicted by the best-fit 2L1S model derived from the ground-based analysis for $\bm{\pi}_{\rm E} = (0,0)$ evaluated at
    the central value of the $VLI$ color-color constraint. The total flux change of the \Sp\ light curve is $\sim 2.0$ instrumental flux unit, which is only a few times the level of systematics seen in other \Sp\ events with observations in the baseline \citep{KB180029,OB170406,OB180799}.}
    \label{lc2}
\end{figure}

\begin{figure}[htb] 
    \centering
    \includegraphics[width=0.65\columnwidth]{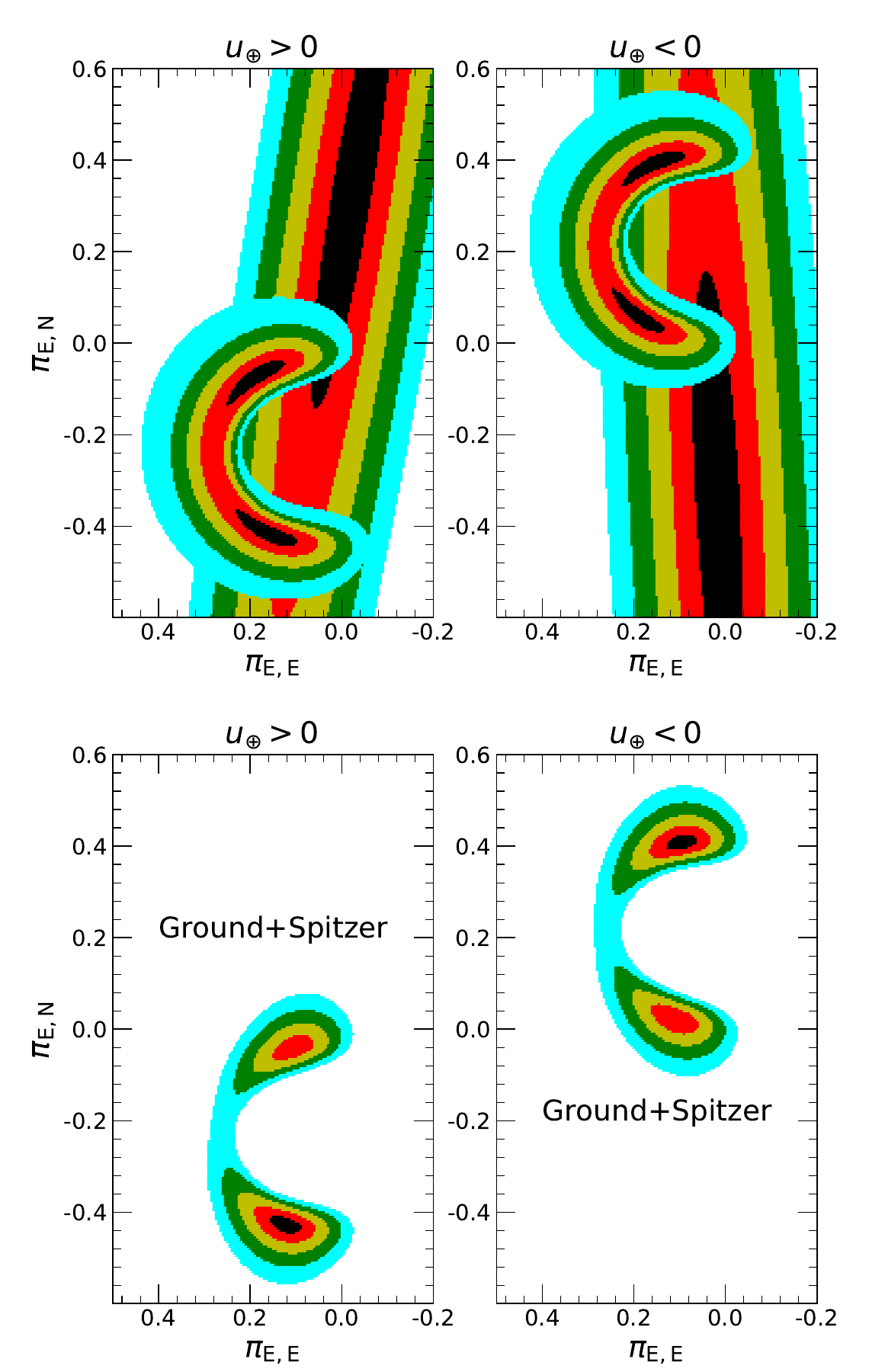}
    \caption{Parallax contours for Ground-ONLY (the elliptical contours in the top panels), {\it Spitzer}-``ONLY'' (the arc-like contours in the top panels) and Ground + \Sp\ (lower panels) parallax analysis. Colors (black, red, yellow, green, cyan) indicate number of $\sigma$ from the minimum (1, 2, 3, 4, 5). For both $u_0 > 0 $ and $u_0 < 0$, the arc-like {\it Spitzer}-``ONLY'' parallax is broken into two discrete minima due to the ``1-D'' constraint of ground-based parallax. The lensing parameters of the four minima are presented in Table \ref{parm2}.}
    \label{pie}
\end{figure}

